\documentclass[final,3p,times,twocolumn]{elsarticle}
\usepackage[hidelinks=true]{hyperref}

\usepackage{graphicx,color,psfrag}
\graphicspath{ {./images/} }
\usepackage{subcaption}

\usepackage{svg}
\usepackage{adjustbox} 

\usepackage[percent]{overpic}
\usepackage{pict2e} 

\usepackage{textcomp,gensymb}

\usepackage{amssymb}
\usepackage[tbtags]{amsmath}

\usepackage{lineno}
\usepackage{geometry}
\usepackage{rotating}
\usepackage{tablefootnote}

\usepackage{nomencl}
\makenomenclature
\usepackage{makeidx}
\makeindex
\journal{Applied Thermal Engineering}

\definecolor{drakgreen}{rgb}{0.0, 0.5, 0.0}

\def \anchoPIV {0.24\textwidth}
\def \tam {\footnotesize}
\def \espacioFigPIVneto {\vspace{0.3cm}}
\def \labelPIVNA{ \scriptsize Average Field}
\def \labelPIVNB{ \scriptsize Instantaneous Field 1}
\def \labelPIVNC{ \scriptsize Instantaneous Field 2}
\def \labelpPIVNAv {30}
\def \labelpPIVNi {20}

\def \RegFlujo {\color{drakgreen} \footnotesize}

\def \espacioFigPIV {\vspace{0.5cm}}
\def \labelPIVA { \tam $\theta=0\degree$}
\def \labelPIVB { \tam $\theta=45\degree$}
\def \labelPIVC { \tam $\theta=90\degree$}
\def \labelPIVD { \tam $\theta=135\degree$}
\def \labelPIVE { \tam $\theta=180\degree$}
\def \labelPIVF { \tam $\theta=225\degree$}
\def \labelPIVG { \tam $\theta=270\degree$}
\def \labelPIVH { \tam $\theta=315\degree$}
\def \labelpPIVAB {40}
\def \labelpPIVCD {35}
\def \labelPIVver {98}
\def \barracolor {\begin{center}
\begin{overpic}[width=7cm]{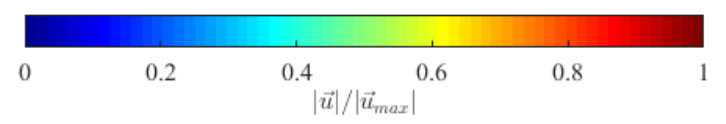}\end{overpic}
\end{center}}
\def \netflowdir {\begin{overpic}[width=2.5cm]{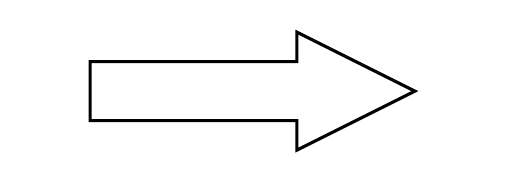} 
\put(30,15){\scriptsize Net flow}
\end{overpic}\\}

\begin{document}

\begin{frontmatter}



\title{Effect of three-orifice baffles orientation on the flow and thermal-hydraulic performance: experimental analysis for net and oscillatory flows}


\author[label1]{J.~Mu\~noz-C\'amara}
\author[label2]{D.~Cresp\'i-Llorens}
\author[label1]{J.P.~Solano}
\author[label2]{P.G.~Vicente}

\address[label1]{Dep. Ing. T\'ermica y de Fluidos. Universidad Polit\'ecnica de Cartagena. Campus Muralla del Mar (30202). Cartagena (Spain).}

\address[label2]{Dep. Ing. Mec\'anica y Energ\'ia. Universidad Miguel Hern\'andez de Elche. Av. de la Universidad s/n (03202). Elche (Spain).}
\address{ Published on Experimental Thermal and Fluid Science on 2021. DOI:\href{https://doi.org/10.1016/j.applthermaleng.2023.121566}{10.1016/j.applthermaleng.2023.121566}. }

\address{ © 2024. Manuscript version under the \href{https://creativecommons.org/licenses/by/4.0/}{CC-BY 4.0 license}.}

\begin{abstract}

Three-orifice baffles equally spaced along a circular tube are investigated as a means for heat transfer enhancement under net, oscillatory and compound flows. An unprecedented, systematic analysis of the relative orientation of consecutive baffles -aligned or opposed- is accomplished to assess the changes induced on the flow structure and their impact on the thermal-hydraulic performance. The results cover the Nusselt number, the net and oscillatory friction factors and the instantaneous velocity fields using PIV in an experimental campaign with a 32 mm tube diameter. The study is conducted in the range of net Reynolds numbers $50 < Re_n < 1000$ and oscillatory Reynolds numbers $0 < Re_{osc}< 750$, for a dimensionless amplitude $x_0/D = 0.5$ and $Pr=65$.  In absence of oscillatory flow, opposed baffles advance the transition to turbulence from $Re_n = 100$ to $50$, increasing the net friction factor (40\%) for $Re_n > 50$ and the Nusselt number (maximum of 27\%) for $Re_n < 150$. When an oscillatory flow is applied, augmentations caused by opposed baffles are only observed for $Re_n < 150$ and $Re_{osc} < 150$. Above $Re_n$, $Re_{osc}>150$, opposed baffles are not recommended for the promotion of heat transfer, owing to friction penalties. However, the chaotic mixing and lack of short-circuiting between baffles observed with flow velocimetry over a wide range of operational conditions point out the interest of this configuration to achieve plug flow.
 
\end{abstract}


\end{frontmatter}


 \setlength{\nomlabelwidth}{2cm}
\begin{thenomenclature}
\nomgroup{A}

\item [{$C$}]  {Propylene glycol weight concentration, \%}
\item[{$c_p$}]{specific heat (J/(kg$\cdot$K))}
\item[{$d$}]{orifice diameter  (m)}
\item[{$D$}]{tube inner diameter  (m)}
\item[{$f$}]{oscillation frequency (Hz)}
\item[{$h$}]{heat transfer coefficient (W/(m$^2\cdot$ K))}
\item[{$I$}]{electric current (A)}
\item[{$k$}]{thermal conductivity (W/(m$\cdot$K))}
\item[{$l$}]{cell length (m)}
\item[{$L_h$}]{heated length (m)}
\item[{$L_p$}]{distance between pressure ports (m)}
\item[{$L_v$}]{baffled tube length in PIV facility (m)}
\item[{$L_{e,v}$}]{entry-outlet length in PIV facility (m)}
\item[{$\dot{m}$}]{mass flow rate (kg/s)}
\item [{$q''$}]  {heat flux, W/m$^2$}
\item [{$Q_{losses}$}]  {heat losses, W}
\item[{$S$}]{open area (-), ${(n \cdot d/D)}^2$}
\item [{$T$}]  {temperature ($^{\circ}$C)}
\item[{$u$}]{velocity vector (m/s)}
\item[{$U$}]{instantaneous bulk flow velocity (m/s), based on $D$}
\item[{$U_n$}]{bulk velocity of the net flow (m/s), based on $D$}
\item[{$V$}]{voltage (V)}
\item[{$x$}]{axial distance from the start of the heated area (m)}
\item [{$x_0$}]  {oscillation amplitude, center to peak (m)}

\item [\underline{Greek symbols}]
\item[{$\beta$}]{coefficient of volumetric thermal expansion (1/K)}{}
\item[{$\mu$}]{dynamic viscosity (kg/(m$\cdot$s))}{}
\item[{$\nu$}]{kinematic viscosity (m$^2$/s)}{}
\item[{$\rho$}]{fluid density (kg/m$^{3}$)}{}
\item[{$\theta$}]{oscillation phase angle ($^{\circ}$)}{}
\item [{$\Delta t$}]  {time step between consecutive images (s)}
\item[{$\Delta p_n$}]{net pressure drop (Pa)}{}
\item[{$\Delta p_{max}$}]{amplitude of the oscillating pressure drop (Pa)}{}

\item [\underline{Subscripts}]
\item[{$b$}]{bulk}
\item[{$in$}]{inlet of the test section}
\item[{$j$}]{section number}
\item[{$k$}]{circumferential position number}
\item[{$wi$}]{inner wall}

\item [\underline{Dimensionless groups}]
\item[{$f_n$}]{net Fanning friction factor, $ \frac{\Delta p_n}{2\rho~U_n^2}~\frac{D}{L_p}$  }
\item[{$f_n$}]{oscillatory Fanning friction factor, $\frac{\Delta p_{max}}{2\rho~(2\pi~f~x_0)^2}~\frac{D}{L_p}$  }
\item[{$Re_n$}]{net Reynolds number, $\rho U_n D / \mu$}
\item[{$Re_{osc}$}]{oscillatory Reynolds number, $\rho (2\pi f x_0) D / \mu$}
\item[{$\Psi$}]{velocity ratio, $Re_{osc}/Re_n$}
\item[{$Pr$}]{Prandtl number, $\mu c_p / k$}
\item[{$Nu$}]{Nusselt number, $h D / k$}
\item[{$Ra^*$}]{Modified Rayleigh number, $g \ \rho \ c_p \ \beta \ D^4 \ q''/(\nu \ k^2)$}

\end{thenomenclature}


\section{Introduction}

The superposition of an oscillatory flow to a net flow in a pipe is a well-known means for heat transfer enhancement \cite{havemann,faghri}, driven by the increase in wall shear stress and the potential change in flow direction. Heat transfer characteristics of pulsating pipe flows have been a focus of interest in a wide variety of applications, ranging from turbomachinery cooling \cite{zheng2020,paniagua,yang2018}, cleaning in place for food industry \cite{augustin2010} or heat sinks for electronics cooling \cite{leong2006}, among others. A comprehensive review of heat transfer enhancement by pulsating flows can be found in \cite{ye_review}. 


When this active technique is combined with artificial tube roughness \cite{li2022,jafari} or with the use of inserts \cite{mackley1990}, the resulting compound enhancement method provides a significant heat transfer augmentation, remarkably when low net flow velocities yield long residence times in the pipe and laminar flow characteristics exist in the absence of pulsation. In particular, the superposition of net and oscillatory flows in tubular baffled reactors, as a means for achieving plug flow characteristics \cite{ni2003mixing}, promotes also high tube-side heat transfer coefficients, allowing to accommodate high heat fluxes in chemical reactions that require simultaneous heating or cooling. Nusselt number correlations for the so-called Oscillatory Baffled Reactors (OBRs) were proposed for the first time by Mackley et al. \cite{mackley1995}, taking into consideration the effect of the net flow velocity and the frequency and amplitude of oscillation using the net Reynolds number and the oscillatory Reynolds number. Single-orifice circular baffles were placed inside the tube, allowing for the oscillatory flow to generate a mechanism of cyclic dispersion of vortices upstream and downstream of the periodically-spaced inserts, that were responsible for the high degree of mixing, reduced axial dispersion and heat transfer enhancement. 

Table \ref{t:review} collects the details (tested range and geometry under study) of previous experimental studies on thermal-hydraulic aspects of OBRs. Only studies with oscillating flow (instead of oscillating baffles) are presented. 

\begin{table*}[ht]
\caption{Experimental studies on Oscillatory Baffled Reactors}
\begin{center}
\begin{tabular}{  c | c | c | c | c | c | c | c| c| c}
    \hline 
     Ref. &  Type of test & $Re_n$ & $Re_{osc}$ & $x_0/D$ & $Pr$ & Geometry & $D (mm)$ & $l$ & $S$ \\
    \hline
    \cite{Mackay1991} & FV, PD & 0 & 300-4000 & 0.07-0.26 & - & MH1 & 25.7 & $1.5D$ & 0.35\\
    \cite{mackley1990} & HT & 100-700 & 200-1600 & 0.12 & 124 & MH1 & 12 & $1.5D$ & 0.34\\
    \cite{mackley1995} & PD, HT & 150-1000 & 300-800 & 0.29-0.57 & 72 & MH1 & 12 & $1.5D$ & 0.60\\
    \cite{stephens} & HT & 0 & 50-1000 & 0.08-0.47 & 102 & MH1 & 24 & $1.5D$ & 0.16\\

    \cite{reis2005} & FV & 0 & 12-560 & 0.05-2.5 & - & SPC & 4.4 & $3D$ & 0.13 \\
    \cite{Zheng} & FV & 0 & 182-300 & 0.08 & - & MH1 & 50 & $1.5D$ & 0.25\\
    \cite{P4G} & HT &  100-1000 & 0-1590 & 0.21 & - & MH1 & 24 & $1.5D$  & -\\    
    \cite{Nogueira} & FV & 0 & 650-1300 & 4 & - & MH3a & 60 & $0.9D$ & 0.24\\
    \cite{Law} & HT & 200-1400 & 0-2770 & 0.23 & 4.4-9  & MH1 & 26.2 & $2D$ & 0.25\\
    \cite{reis2018} & HT & 10-55 & 0-200 & 0.1-0.4 & 5.4 & SPC & 5 & $2.6D$  & 0.16\\
    \cite{MunozMH3} & PD, HT & 10-600 & 10-440 & 0.5 & 190-470 & MH3a & 32 & $0.87D$ & 0.25\\
    \cite{MunozMH1osc} & FV, HT  & 20-65 & 0-650 & 0.5 & 150 & MH1 & 32 & $1.5D$ & 0.25\\
    \cite{MunozPower} & PD & 0 & 10-1000 & 0.14-0.50 & - & MH1, MH3a & 32 & $1.5D$ & 0.25\\
    \cite{baird_ramarao} & PD & 0 & 4·10$^3$-36·10$^3$ & 0.025-0.05 & - & MH3a, MH3o & 19.4 & $0.25D$ & 0.169\\
    
    \hline
    \multicolumn{10}{l}{\small Type of test: HT=Heat transfer, PD=Pressure drop, FV=Flow visualization} \\
    \multicolumn{10}{l}{\small Geometry: MH1: one-orifice baffles, MH3a= three-orifice aligned baffles, MH3o= three-orifice opposed baffles} \\
    \multicolumn{10}{l}{\small  SPC=Smooth Periodic Constrictions} \\
\end{tabular}
\end{center}
\label{t:review}
\end{table*}


The scale-up of single-orifice OBRs, however, showed a reduction of intensity and length-scale of mixing. To overcome this problem, Smith and Mackley \cite{smith2006} suggested the use of multi-orifice baffles, that would mimic the many smaller diameter tubes in parallel proposed by Ni \cite{ni1994} as a solution for the scaling-up of single-orifice OBRs. This approach would be rather similar to a reciprocating plate column \cite{karr}, but oscillating the fluid and not the baffles.


One distinctive feature of the multiorifice designs is that they offer the option to rotate alternatively consecutive baffles, in order to break the strict flow periodicity. This involves a modification of the flow structures, in comparison with aligned multiorifice baffles. Heat transfer performance, energy dissipation or radial mixing might also be affected by this misalignment. However, while aligned baffles have been generally characterized (flow behaviour \cite{smith2000},  mixing \cite{Gonzalez}, heat transfer and pressure drop \cite{MunozMH3}), the effect of opposed multiorifice baffles has rarely been treated in the open literature \cite{baird_ramarao}. In their publication, Baird and Rao \cite{baird_ramarao} studied the effect of the misalignment of multiorifice plates in oscillatory baffled columns on power dissipation. They found a significant influence of the misalignment for a distance between plates $l=50$ mm and oscillating amplitude $x_0=10$ mm, with $D=190$ mm. The differences, however, were negligible for $x_0=5$ mm, as well as for $l=100$ mm and both amplitudes. The authors did not provide any measurement uncertainty analysis, which limits the significance of the results. The absence of flow oscillation also limited the conclusions of this work to the corpus of oscillatory baffled reactors. To the best of our knowledge, no further investigations on the misalignment of multiorifice plates in OBRs have been reported so far in the open literature.


The lack of strict periodicity of the baffles' orientation in OBRs has been conversely studied from the perspective of different axial baffle arrangements. In their search for the rupture of the shortcut between consecutive baffle orifices, Mazubert et al. \cite{mazubert} introduced a central baffle between single-orifice baffles. The authors reported a significant increase in pressure drop and shear strain rate, but neither the axial dispersion nor the radial mixing were improved. Zhang et al. \cite{zhang2022} evaluated different chambers connecting consecutive cells in oscillatory baffled crystallizers, with the aim of breaking the shortcut. They demonstrated the positive impact of these designs on the particle suspension performance of the reactor.

On the basis of this literature review, the impact of the misalignment of consecutive orifice baffles on the fluid mechanics and heat transfer performance of OBRs is still unclear. Most of the designs rely on the generation of jets across the orifices that yield vortex dispersion and further dissipation of the accelerated flow along each cell tank. Depending on the amplitude of the oscillation, a shortcut phenomenon might occur when jets cross aligned baffle orifices. The misalignment of consecutive orifice baffles, however, might involve different flow pattern mechanisms, whose description and influence on the friction and heat transfer characteristics of the flow have not been analysed. 

In this work, a global approach to these open questions is applied to a tube with three-orifice baffles orientation with aligned and opposed configurations working under net, oscillatory and compound flow conditions. To this aim, three main measurement techniques are used in both a thermal-hydraulic and a visualization rig: (1) heat transfer tests to obtain the Nusselt number under uniform heat flux conditions and a Prandtl number of 65, (2) pressure drop tests under isothermal conditions to determine the net and oscillatory Fanning friction factors, and (3) PIV tests to derive the instantaneous velocity fields. The results are presented and discussed in different sections according to the type of flow: net, oscillatory or compound. The dimensionless amplitude tested was the same for all the cases, $x_0/D=0.5$, covering a wide range of Reynolds numbers: $50<Re_n<1000$ and $0<Re_{osc}<750$. The outcomes of this investigation allow us to establish the operational regions where opposed baffles can provide a significant enhancement over the aligned baffles.

\section{Experimental methodology}

\subsection{Baffles tested}

\begin{figure}[htbp]
\centering
\includegraphics[width=8cm]{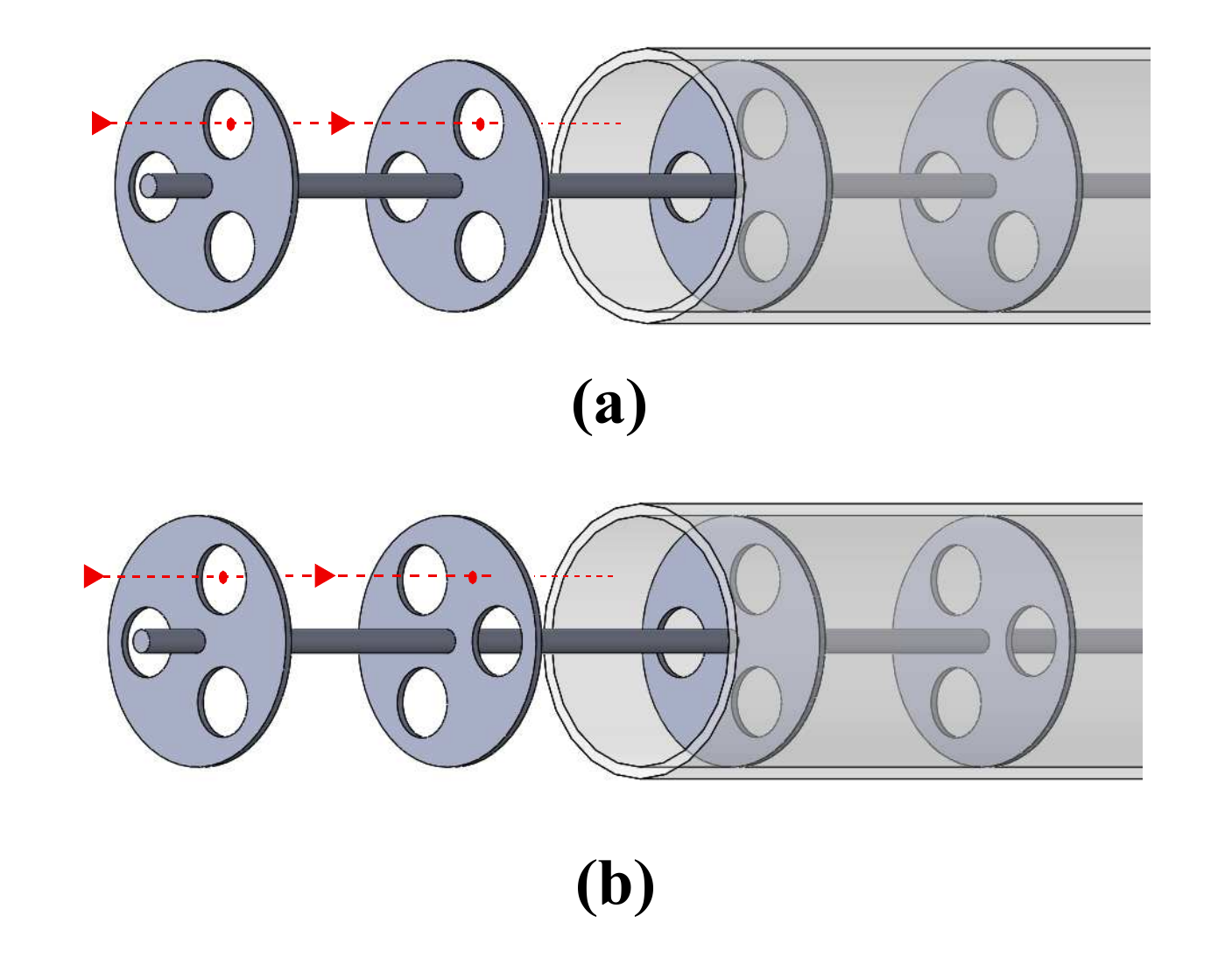}
\caption{Three-orifice baffles configurations tested: (a) aligned and (b) opposed. \label{f:baffle}}
\end{figure}

The baffles selected for the present investigation are depicted in Fig.~\ref{f:baffle} and will be abbreviated as MH3 baffles from this point forward. Three-circular orifices, $n_o=3$, separated 120$^{\circ}$ provide the same open cross-section area than the standard orifice baffle, $S=0.25$ (which was proposed as the optimum value by Ni et al. \cite{ni2003mixing} from the results of mixing times for different operating conditions). The distance between baffles, $l$, is shorter if compared to one orifice baffles where $l=1.5\cdot D$, allowing for the appropriate dispersion of vortices along smaller cells. Following the recommendation from González-Juárez et al. \cite{Gonzalez}, the ratio $l=1.5\cdot D_{eq}$ is kept constant, where $D_{eq}=D/\sqrt{n_o}$. 

The baffles are made of PEEK plastic, to avoid electrical and thermal conduction from the tube wall, with thickness $e=1$ mm.

Consecutive baffles present the same orientation with respect to the tube axis in the aligned baffle arrangement (Fig.~\ref{f:baffle}a). Conversely, the opposed baffle arrangement (Fig.~\ref{f:baffle}b) is built with a rotation of 60$^{\circ}$ between consecutive baffles. 

\subsection{PIV facility}

\begin{figure*}[htbp]
\centering
\subcaptionbox{PIV facility setup: (1) Centrifugal pump, (2) flow control valve, (3) input flow temperature probe B 1/10 DIN PT100, (4) test section with insert baffles and compensation box, (5) output flow temperature probe B 1/10 DIN PT100, (6) manual valve, (7) Coriolis flowmeter, (8) plate heat exchanger, (9) chiller, (10) reservoir tank, (11) electric heater with temperature control system, (12) PIV high speed camera equipped with a MACRO lens, (13) PIV laser, (14) motor-gear assembly, (15) rod-crank system, (16) double effect piston and (17) magnetostrictive position sensor.\label{f:PIVSetup}}[14cm]{ \includegraphics[width=12cm]{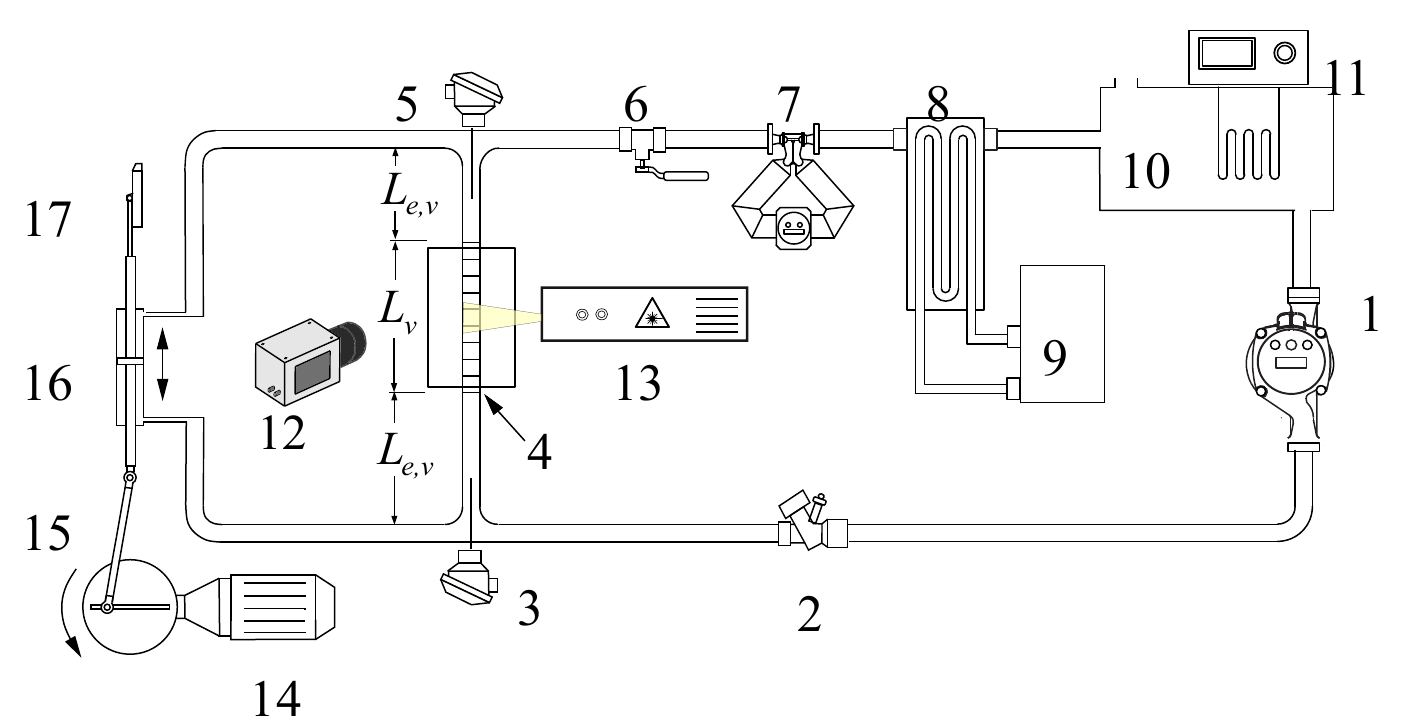}}
    \\
\subcaptionbox{Thermohydraulic facility setup. (1) reservoir yank, (2) gear pump, (3) Coriolis flowmeter, (4 \& 12) manual valves, (5 \& 11) input/output temperature probes B 1/10 DIN PT100, (6) test section with insert baffles, (7) type T thermocouples set, (8) SMAR LD-301 static differential pressure sensors, (9) Kistler 4264A dynamic differential pressure sensors, (10) autotransformer, (13) agitator, (14) double effect hydraulic piston, (15)  rod-crank assembly, (16) gear-motor group, (17) magnetostrictive position sensor\label{f:THSetup}}[14cm]%
{ \includegraphics[width=14cm]{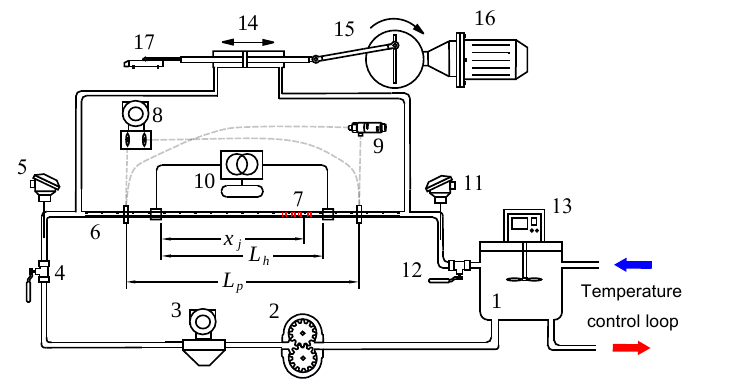}}
\caption{Experimental facilities.\label{f:facilities}}
 \end{figure*}

The facility depicted in Fig.~\ref{f:facilities}a was built for visualization purposes. The test section is made of acrylic tube with equally spaced inserted baffles. A total number of 12 baffles  ensures the spatial periodicity of the flow field in its middle section. The test section is $L_v=331$ mm long and complies with the dimensions in Fig.~\ref{f:tps} with a wall thickness of 3 mm. The flow is visualized in the space between consecutive baffles. Around the visualization area, a flat-sided acrylic box, filled with the same working fluid, is placed to reduce optical distortion. Water is used as working fluid.

The temperature of the working fluid is adjusted using an electric heater~(11) and a chiller~(9) connected through a plate heat exchanger~(8). A centrifugal pump~(1) circulates the net flow from the reservoir tank~(10) to the test section in a closed loop. The superposition of an oscillatory flow is achieved by the connection of a double-effect hydraulic cylinder~(16) to the test section. The use of a slider-crank mechanism~(15) allows for its activation, generating a sinusoidal linear displacement with frequencies between 0.12~Hz and 1.2~Hz. A flow control valve~(2) is used to ensure a stable net flow. The position of the hydraulic cylinder~(16) is measured by a magnetostrictive position sensor~(17).

The flow is seeded with  57~$\mu$m  polyamide particles with a density of 1051~kg/m$^3$. A 1~mm thick laser light sheet of 808~nm wavelength illuminates a vertical plane of the tube. Consecutive images of the seeded flow are captured by a 1280$\times$1024 pix$^2$ CMOS IDT MotionScope M3 high speed camera. The spatial resolution of the images is around 0.031 mm. PIV is carried out by using PIVlab v2.31 for Matlab~\cite{PIVlab}.

After PIV image pre-processing (background removal, histogram equalization, intensity high-pass and intensity capping), 97.7\% of the velocity vectors are found valid, $PPR>2$ (peak-to-peak ratio) for the net flow tests. Image processing is carried out by the adaptive FFT (Fast Fourier Transform) cross correlation algorithm in four steps, with interrogation area sizes: 192\(\times\)192~pix$^2$, 128\(\times\)128~pix$^2$, 96\(\times\)96~pix$^2$ and 88\(\times\)88~pix$^2$ (all of them with a 50\% overlap), leading to a spatial resolution of the velocity field equal to 1.35 mm. Post-processing includes the application of a global filter (based on the manual selection of the points that deviate significantly from the main cloud of points, when $x$ and $y$ velocity components are represented) and two local filters (based on the standard deviation and the median of the neighbour windows velocities).

Given the dissimilar velocities found both in the space and time domains, a time step between snapshots that ensured $PPR>2$ for at least 85\% of the velocity vectors at $\theta=0^\circ$ (maximum average velocity) was chosen. 

Table~\ref{table_piv} summarizes the PIV tests carried out for both geometries and the time step between consecutive images that was used. This nomenclature (N1, N2, etc.) is also used along the paper to represent the tested points in the pressured drop and heat transfer results.

\begin{table*}[ht]
\caption{PIV tests data}
\begin{center}
\begin{tabular}{  c | c | c | c | c | c | c }
    \hline 
     Test & $\dot{m}$ (kg/h) & $x_0$ (mm) & $f$ (Hz) & $Re_n$ & $Re_{osc}$ & $\Delta t$ (ms)\\
    \hline
    N1 & 10 & - & 0 & 115 & 0 & 20 \\
    N2 & 20 & - & 0 & 230 & 0 & 10 \\
    N3 & 30 & - & 0 & 460 & 0 & 5 \\
    \hline
    O1 & 0 & 16 & 0.056 & 0 & 190 & 20 \\
    O2 & 0 & 16 & 0.112 & 0 & 380 & 6.7 \\
    O3 & 0 & 16 & 0.165 & 0 & 570 & 5 \\
    \hline
    C1 & 10 & 16 & 0.056 & 115 & 190 & 20 \\
    C2 & 10 & 16 & 0.112 & 115 & 380 & 6.7 \\
    C3 & 10 & 16 & 0.165 & 115 & 570 & 5 \\
    \hline
\end{tabular}
\end{center}
\label{table_piv}
\end{table*}

\subsection{Thermohydraulic facility}

Fig.~\ref{f:THSetup} shows the thermohydraulic facility, where three different types of tests have been performed. The test section consists of a 2 meter long 316 stainless steel tube, with 32 mm i.d. and 1.5 mm wall thickness, where the baffles are introduced. The main tank (1) includes an electrical heater and mixer (13) that ensures a homogeneous mixture and heats the fluid when it is required. An external loop can keep constant the tank temperature in the range 15-40 $^{\circ}$C.

Table \ref{table_th} shows the values of the variables that have been modified to cover the tested ranges. The ranges tested were selected to reproduce conditions.

\begin{table*}[ht]
\caption{Thermohydraulic tests data (for both configurations)}
\begin{center}
\begin{tabular}{  c | c | c | c | c | c | c }
    \hline 
    Type of test & $C$ (\%) & $T_b$  ($^{\circ}$C) & $\dot{m}$ (kg/h) & $x_0$ (mm) & $f$ (Hz) & $q''$ (W/m$^2$) \\
    \hline
    Net pressure drop & 50, 80 & 22 & 30-500 & - & 0 & 0 \\
    Oscillatory pressure drop & 50, 80 & 22 & 0 & 16 & 0.22-2.2 & 0 \\
    Heat transfer & 50 & 30* & 30-500 & 16 & 0, 0.26, 0.66, 0.98, 1.30 & 5500-9000 \\
    \hline
    \multicolumn{7}{l}{\small *Bulk temperature measured at the first thermocouple section, $T_{b,1}$.} \\
\end{tabular}
\end{center}
\label{table_th}
\end{table*}

The flow rate range has been selected to reproduce the net Reynolds number range $50<Re_n<1000$ that: (1) covers the expected critical net Reynolds number ($\approx$100 \cite{MunozMH3}) for the aligned baffles, so the differences between both configurations can be appreciated more clearly, and (2) there are potential applications, because a smooth tube would produce intense mixing at higher Reynolds numbers ($Re>4000$).

The particularities for each type of test are described below.

\subsubsection{Net pressure drop tests}

The net pressure drop tests are carried out with valves (4) and (12) open, the oscillatory system (14-16) and transformer (10) are turned off, so there is only net flow (provided by the main pump (2)) and no heating is applied. These tests are performed under isothermal conditions at room temperature.

The net pressure drop is measured by a differential pressure transducer, model LD301 (8). Two different ranges are used for this study: 0-50 mbar and 0-500 mbar. The distance between pressure ports in the test section is 1295 mm, with 46 baffles located between them.

The fluid properties are evaluated at the mean temperature in the test section (PT100s (5) and (11)). Two propylene glycol-water mixtures, 50\% and 80\% propylene glycol, are used to widen the range of net Reynolds numbers covered and, at the system, check the repeatability of the measurements.

For each net Reynolds number tested, 100 samples are taken by the datalogger (model Agilent 34972A) at a frequency of 1 Hz.

\subsubsection{Oscillatory pressure drop tests}

For the oscillatory pressure drop tests, the main pump (2) is turned off and valves (4) and (12) are closed. The system is previously pressurized to avoid the potential risk of cavitation in the test section. Again, the transformer is not connected, and the tests are performed at room temperature under isothermal conditions.

The flow takes place in the oscillatory loop, provided by the motion of a double-effect hydraulic cylinder (14). A crank and slider mechanism (15) ensures that if follows a quasi-sinusoidal motion when it is connected to the motor (16). A magnestroctive sensor (17) measures the instantaneous position of the cylinder rod.

The instantaneous pressure drop between the pressure ports (the same used for the net pressure drop tests) is acquired by piezoresistive differential pressure sensors, model KISTLER 4264A, ranges $\pm$ 100 mbar and $\pm$ 1000 mbar. These sensors have a response frequency of 2000 Hz, enough to resolve the pressure drop signal for an oscillating cycle.

The high frequency signals (pressure drop and position) are acquired by the acquisition card USB-6001 (National instruments) at a sample rate of 2.8 kHz for each channel. At least 20 oscillation cycles have been obtained for each oscillatory Reynolds number tested.

The fluid properties are evaluated at the mean temperature in the test section (from (5) and (11)).

To characterize the oscillatory pressure drop, the oscillatory Fanning friction factor is calculated. It is based on the common definition of the Fanning friction factor, but using the maximum oscillatory velocity, $2\pi~f~x_0$, and the pressured drop amplitude, $\Delta p_{max}$, as the characteristic velocity and pressure drop, respectively.

\begin{equation}
f_{osc}=\frac{\Delta p_{max}}{2\rho~(2\pi~f~x_0)^2}~\frac{D}{L_p}
\label{eq_fosc}
\end{equation}

$\Delta p_{max}$, is obtained by a nonlinear statistical fitting (of the real pressured drop signal) to a sine wave. The value of the fitted amplitude is taken as $\Delta p_{max}$.

\subsubsection{Heat transfer tests}

The heat transfer tests are performed for net flow conditions and compound conditions (net + oscillatory flow). Thus, valves (4) and (12) are open and the heating by Joule effect applied (10). The only difference between the tests for net flow and compound flow is the state of the oscillatory system (14-16), turned off or not, respectively.

The local Nusselt number at a given axial position is evaluated from:

\begin{equation}
    Nu_j = \frac{V \cdot I-Q_{losses}}{\pi L_h (T_{wi,j}-T_{b,j})} \cdot \frac{1}{k_j}
\end{equation}

$V$ and $I$ are the voltage and intensity applied by the transformer, their product corresponds to the raw heat applied to the tube. The voltage is measured directly by the datalogger, while the intensity is measured by a hall effect sensor. $L_h$ is the heated section length, which corresponds, for this case, to the distance between the two electrodes that connect the test section to the transformer, $L_h=1420$ mm. $k_j$ is the thermal conductivity of the test fluid evaluated at the bulk temperature at the axial position $j$. It should be noted that the inner pipe wall is considered as the heat transfer surface area, due to the low thermal conductivity of the baffles.

$T_{b,j}$ is the fluid bulk temperature at the axial position $j$, which can be calculated as:

\begin{equation}
    T_{b,j}=T_{b,in}+\frac{V \cdot I-Q_{losses}}{\dot{m}\cdot c_p} \cdot \frac{x_j}{L_h}
\end{equation}

Where $T_{b,in}$ is the inlet fluid temperature (from PT100 (5)), $\dot{m}$ is the mass flow rate (Coriolis flow meter (3)), and $c_p$ is the specific heat of the fluid at the inlet temperature. $x_j$ is the distance between the axial position $j$ and the start of the heated section.

\begin{figure}[htbp]
\centering
\subcaptionbox{Aligned baffle arrangement.\label{f:MH3aligned}}[7cm]{ \includegraphics[width=7cm]{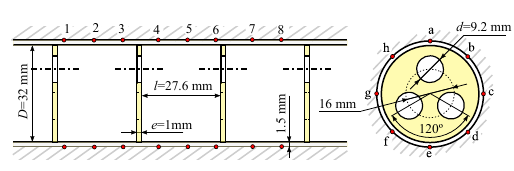}}
\subcaptionbox{Misaligned baffle arrangement.\label{f:MH3misaligned}}[7cm]{ \includegraphics[width=7cm]{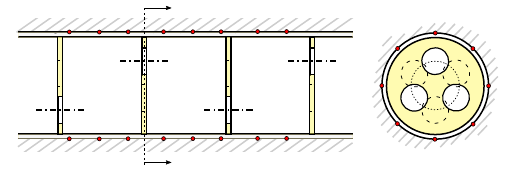}}
\caption{Dimensions of the baffles and thermocouple disposition (red dots) for aligned (a) and opposed (b) baffles.\label{f:tps}}
\end{figure}

The inner wall temperature cannot be measured directly, so the outer wall temperature is taken instead by type T thermocouples (7).For each axial position $j$, 8 thermocouples equally distributed are placed on the outer wall of the tube as depicted in Fig.~\ref{f:tps} (ensuring an accurate measurement when flow stratification is present). The average of these 8 measurements is taken as the outer wall temperature. To derive the inner wall temperature from the outer, a 1D discrete model is used to solve the heat conduction with internal heat generation in the tube.

The heat losses, $Q_{losses}$, are estimated from the averaged outer wall temperature, by considering the thermal conductivity and thickness of the thermal insulation layer and the outer heat transfer coefficient (natural convection) from well-known correlations for a horizontal cylinder \cite{Churchill}. The tube was thermally insulated with a layer of rockwool (thermal conductivity of 0.035 W/(m·K)) with a thickness of 50 mm. The outer surface was covered with a 5 mm thick multilayer insulation (aluminium and polyethylene) of low emissivity (0.06).

These calculations are repeated for 8 axial sections to catch the variability of the Nusselt number along one cell (distance between baffles), which is due to the fact that the baffles prevent the flow development. Because one cell length is too short to place 8 measurement sections (64 thermocouples in total), the periodicity of the flow is considered to place the measuring section in 8 different cell positions (relative to the previous baffle) and different cells.

For all the signals, 30 samples for each test are acquired by two dataloggers (model Agilent 34970A and 34972A), with an acquisition rate of 0.1 Hz.

\subsection{Fluid properties}

The required fluid properties (density, dynamic viscosity, specific heat and thermal conductivity) are calculated by interpolation on the tabulated data of propylene-glycol and water mixtures included in ASHRAE’s 2001 fundamentals book \cite{ASHRAE}, as a function of the temperature and the propylene-glycol concentration.

Because the propylene-glycol concentration is sensitive to previous tests (dead zones in the hydraulic circuit) and the humidity in the lab (propylene-glycol is hygroscopic), it is important to consider a method to periodically measure the real concentration. To do so, a Cannon-Fenske viscometer is used to measure the kinematic viscosity in a thermostatic bath at a known temperature (around 30 $\degree$C), then the concentration that better fits the measured viscosity is considered. The high sensitivity of the fluid viscosity with the concentration provides a low uncertainty.

\subsection{Uncertainty analysis}

The uncertainty analysis follows the methodology included in \cite{Dunn_uncer}. As a sum up, the uncertainty of each direct measurement (e.g., net pressure drop) includes the bias component (obtained from the sensor manufacturer, collected in Table~\ref{t:sensors}) and the random component (derived from the standard deviation of the corresponding measurements of a variable); then, error propagation is applied to calculated the global uncertainty of the variables of interest (e.g., net friction factor).

\begin{table*}[htbp]
\caption{Measurement uncertainties \label{t:sensors}}
\centering
	\begin{tabular}{ll}
		\hline
		Measurement & Uncertainty \\ \hline
            Mixture concentration,  $C$ & 0.7\% \\ 

            Tube diameter, $D$ & 0.1\% measure \\ 
		Heat transfer section, $L_h$ & 0.01 m \\ 
		Thermocouples position, $x_j$ & 0.005 m \\ 
		Pressure test length, $L_p$ & 0.005 m \\
  
            Mass flow rate, $\dot{m}$ & 0.2\% measure \\ 
            Net pressure drop, $\Delta p_n$ & 0.1\% full scale (50 or 500 mbar)\\

            Frequency, $f$ & 0.004 Hz \\
            Oscillation amplitude, $x_0$ & 0.3 mm \\
            Oscillatory pressure drop, $\Delta p_{max}$ & 0.2\% full scale (100 mbar or 1 bar)\\
            
  		Inlet temperature, $T_{b,in}$ & $0.1 \cdot (0.03+0.005 \cdot T(^\circ C))$ \\ 
		Outer wall temperature, $T_{we}$ & 1.0 $^\circ$C \\ 
		Voltage, $V$ & 0.06\% measure + 0.04\% full scale (10 V)\\ 
		Intensity, $I$ & 2\% full scale (600 A) \\ 
            Heat losses,  $Q_{losses}$ & 50\% measure\\  
		 \hline
	\end{tabular}
\end{table*}

As can be observed, the fluid properties uncertainties are not included in Table~\ref{t:sensors}, because these are not a variable directly measured, but derived from the concentration and temperature of the fluid. Consequently, they are calculated for each tested by applying the error propagation (from the uncertainties of the fluid temperature and concentration). Because the properties are derived from interpolation on tabulated data, the partial derivatives to apply the error propagation formula were obtained numerically.

The uncertainty for each test for a 95\% confidence interval is included directly in all the figures throughout this work. Table~\ref{t:uncer} shows the maximum and average uncertainty for the different results of interest.

\begin{table}[htbp]
\caption{Uncertainties of the relevant variables\label{t:uncer}}
	\begin{tabular}{lll}
		\hline
		Variable & Max. uncertainty & Mean uncertainty\\ \hline
		$Re_{n}$ & 5.8\% & 2.2\%\\ 			
	    $Re_{osc}$ & 5.7\%  & 3.8\%\\ 
  		$x_0/D$ & 1.8\% & 1.8\%\\ 	
		$Pr$ & 3.4\% & 1.9\%\\ 
            $Ra*$ & 4.7\% & 3.5\%\\ 
		$f_n$ & 15.1 & 4.2\%\\ 
		$f_{osc}$ & 16.0\% & 5.7\%\\ 
		$Nu$ & 10.2\% & 6.9\%\\  \hline
	\end{tabular}
\end{table}

\section{Results}

\subsection{Net flow}

\begin{figure}[htbp]
\centering
\includegraphics[width=8cm]{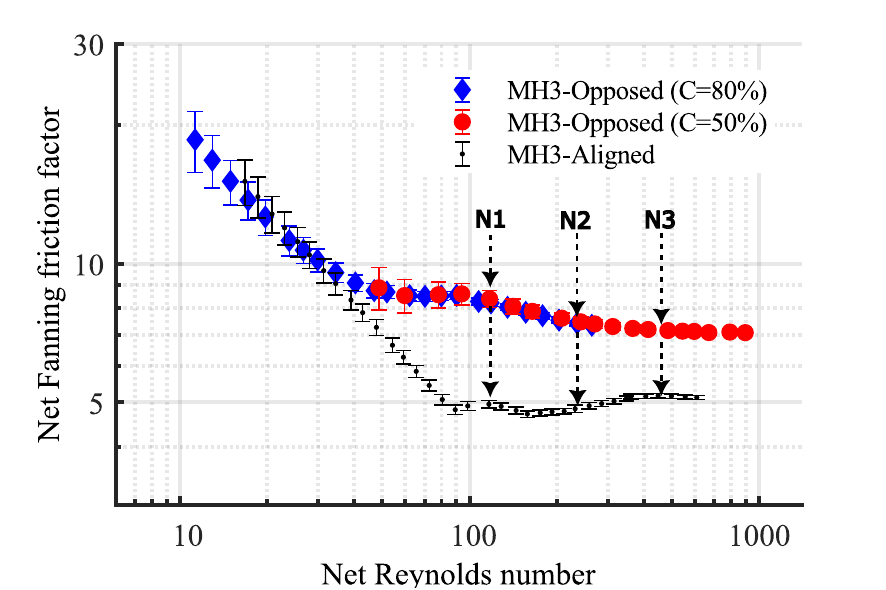}
\caption{Net Fanning friction factor vs net Reynolds number for both MH3 baffle types. \label{f:f_net}}
\end{figure}

\paragraph{Pressure drop}
Fig.~\ref{f:f_net} shows the Fanning friction factor under net flow conditions for both the aligned and the opposed three orifice baffles. The results for the opposed baffles are obtained for two different propylene glycol-water mixtures to increase the range of net Reynolds numbers tested and to prove the consistence of the results where the ranges for both mixtures overlap ($Re_n=50-300$).

At low net Reynolds numbers, $Re_n<40$, both configurations display the same trend and differences below the experimental uncertainty. However, the opposed baffles experience a sharp change in the curve slope at $Re_n \approx 40$, indicating the end of the laminar flow regime, while the aligned baffles show a similar behaviour at $Re_n \approx 90$. Thus, both configurations imply a significant reduction in the critical net Reynolds number when compared to one-orifice baffles, where $Re_n \approx 165$ \cite{MunozMH1net}.

The transition occurs smoothly when compared to a smooth tube, and the trends are different enough to prevent the use of one of the available methods to delimit the different flow regimes \cite{Meyer_regions}. 

Beyond the laminar flow region, the friction factor is significantly higher for the opposed baffles, a maximum of 60\% for N1 ($Re_n=115$), and around 40\% in the region where the friction factor for both configurations seems to become stable (N3).

\paragraph{Velocity field}
The velocity field in pure net flow conditions has been obtained in a symmetry plane of the tube at $Re_n=115$ (N1) and $Re_n=460$ (N3) for both configurations. The analysis in these operational conditions is aimed at helping to identify the smooth transition to turbulence observed in   pressure drop results (Fig.~\ref{f:f_net}).

\begin{figure*}[htbp]
\subcaptionbox{Aligned baffle arrangement.\label{f:PIV_N1_ALI}}[\textwidth]{ \begin{overpic}[width=\anchoPIV]{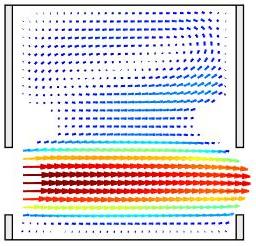} 
\put (\labelpPIVNAv,\labelPIVver) {\labelPIVNA}

\put(55,-5){\RegFlujo \vector(0,1){20}~ Jet}

\put(0,-5){\RegFlujo Recirculation \vector(0,3){60}}
\put(7.5,47){\RegFlujo
\polygon(0,0)(82.5,0)(82.5,44)(0,44)}

\end{overpic} \begin{overpic}[width=\anchoPIV]{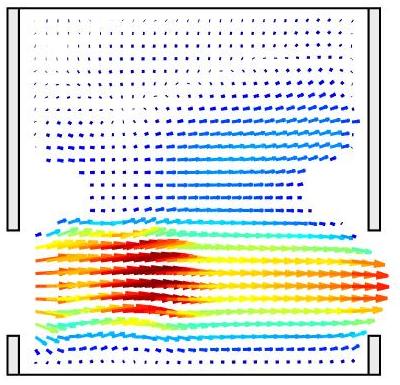} 
\put (\labelpPIVNi,\labelPIVver) {\labelPIVNB}
\end{overpic} \begin{overpic}[width=\anchoPIV]{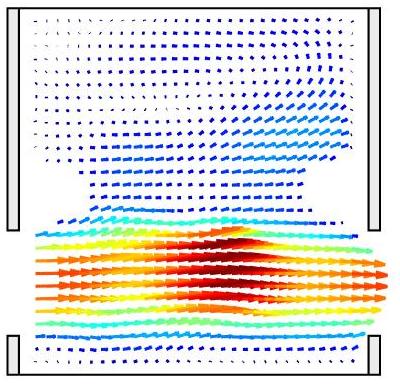}
\put (\labelpPIVNi,\labelPIVver) {\labelPIVNC}
\end{overpic}}\\

\espacioFigPIVneto

\subcaptionbox{Opposed baffle arrangement. \label{f:PIV_N1_DES}}[\textwidth]{ \begin{overpic}[width=\anchoPIV]{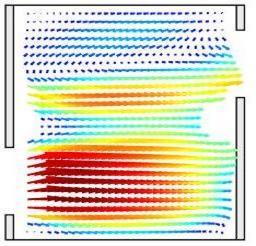} 
\put (\labelpPIVNAv,\labelPIVver) {\labelPIVNA}
\put(95,75){\RegFlujo I }
\put(75,55){\RegFlujo
\polygon(0,0)(25,0)(25,30)(0,30)}
\end{overpic} \begin{overpic}[width=\anchoPIV]{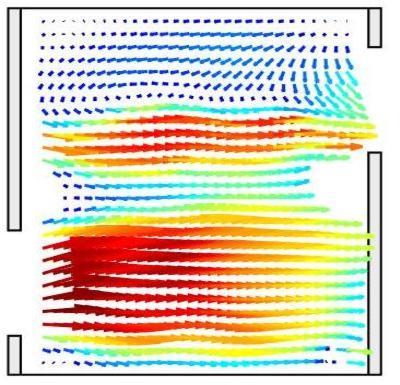} 
\put (\labelpPIVNi,\labelPIVver) {\labelPIVNB}
\end{overpic} \begin{overpic}[width=\anchoPIV]{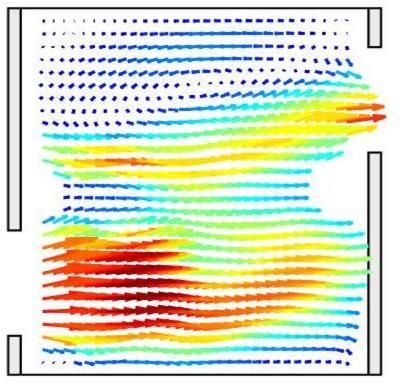}
\put (\labelpPIVNi,\labelPIVver) {\labelPIVNC}
\end{overpic}}\\
\barracolor
\caption{PIV velocity fields at $Re_n=115$ for the geometries under study. Experiment N1 (Table~\ref{table_piv}). \label{f:PIV_N1}}
\end{figure*}
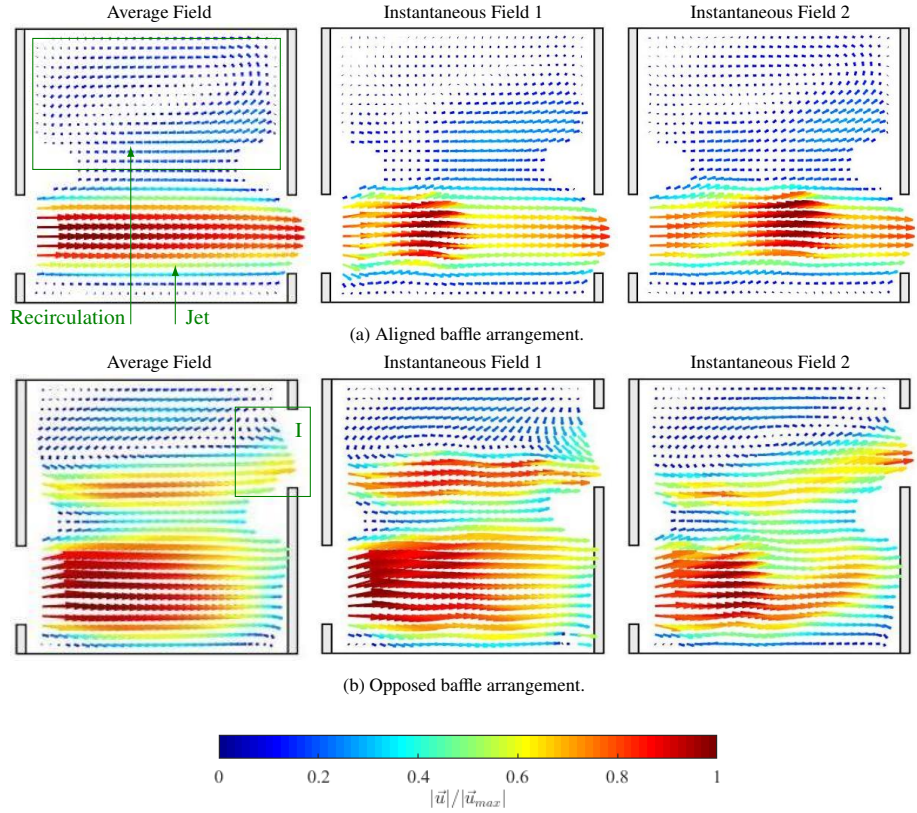

\begin{figure*}[htbp]
\subcaptionbox{Aligned baffle arrangement. \label{f:PIV_N3_ALI}}[\textwidth]{ \begin{overpic}[width=\anchoPIV]{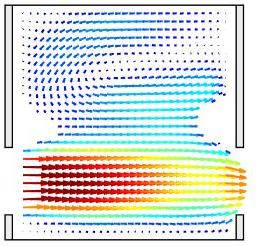} 
\put (\labelpPIVNAv,\labelPIVver) {\labelPIVNA}
 \put(80,-5){\RegFlujo \vector(0,1){20} I}
\put(-7,58){\RegFlujo II}
\put(0,60){\RegFlujo \vector(3,0){60}}
\put(0,62){\RegFlujo \vector(6,3){20}}

\end{overpic} \begin{overpic}[width=\anchoPIV]{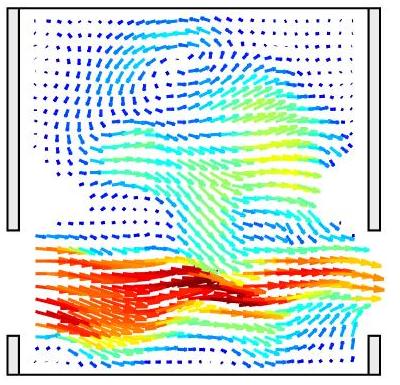} 
\put (\labelpPIVNi,\labelPIVver) {\labelPIVNB}
\end{overpic} \begin{overpic}[width=\anchoPIV]{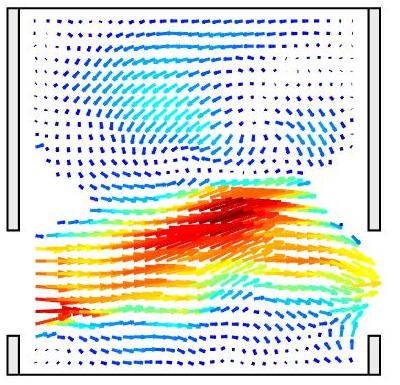}
\put (\labelpPIVNi,\labelPIVver) {\labelPIVNC}
\end{overpic}}\\

\espacioFigPIVneto

\subcaptionbox{Opposed baffle arrangement. \label{f:PIV_N3_DES}}[\textwidth]{ \begin{overpic}[width=\anchoPIV]{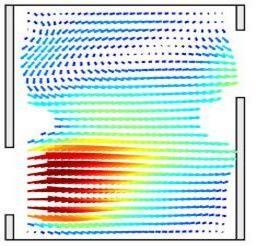} 
\put (\labelpPIVNAv,\labelPIVver) {\labelPIVNA}
\put(35,-5){\RegFlujo \vector(0,1){15}~ Shorter jet} \end{overpic} \begin{overpic}[width=\anchoPIV]{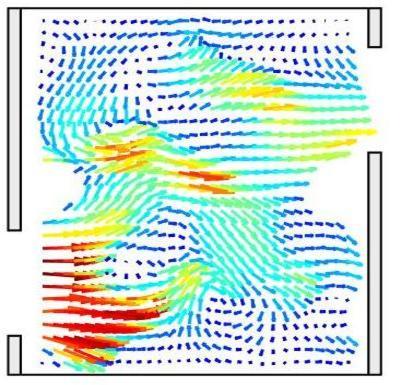} 
\put (\labelpPIVNi,\labelPIVver) {\labelPIVNB}
\end{overpic} \begin{overpic}[width=\anchoPIV]{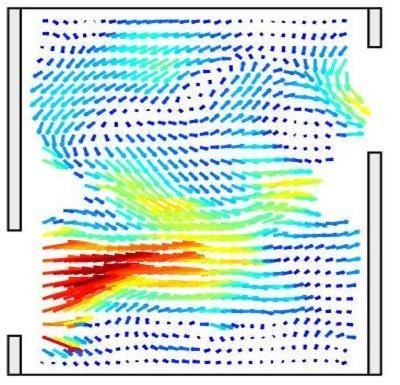}
\put (\labelpPIVNi,\labelPIVver) {\labelPIVNC}
\end{overpic}}
\barracolor
\caption{PIV velocity fields at $Re_n=460$ for the geometries under study. Experiment N3 (Table~\ref{table_piv}).\label{f:PIV_N3}}
\end{figure*}

As shown in Fig.~\ref{f:PIV_N1_ALI} and Fig.~\ref{f:PIV_N3_ALI}, for $Re_n =115$ and $Re_n= 460$, respectively,  the average flow pattern for the aligned geometry is dominated by a jet connecting the orifices of consecutive baffles 
and a recirculation bubble that extends along the upper half of the cell. (Fig.~\ref{f:PIV_N3_ALI}). 
Increasing the net Reynolds number results in lower velocities in the final part of the jet (Fig.~\ref{f:PIV_N3_ALI}-I), presumably due to higher momentum diffusivity, and higher local velocities in the recirculation bubble (Fig.~\ref{f:PIV_N3_ALI}-II). 

The average velocity field depicted in Fig.~\ref{f:PIV_N1_DES} and Fig.~\ref{f:PIV_N3_DES} for the opposed baffles' arrangement allows us to observe a similar jet structure downstream of the baffles. However, the misalignment of the baffle located immediately downstream prevents the further development of the jet. The recirculation in the upper part of the cell is affected by the three-dimensional flow structures originated by the blockage encountered by the jet. Local axial velocities in the visualization plane are higher than radial velocities. The evacuation of the flow through the upper orifice of the baffle downstream (Fig.~\ref{f:PIV_N1_DES}-I) is also detected. The increase of the net Reynolds number causes the jet to dissipate faster, being significantly shorter for $Re_n=460$ than for $Re_n=115$. 
As shown in pressure drop experiments for both configurations, the flow presents a smooth transition to turbulence. A good measure for the flow instability is to compare average and instantaneous velocity fields. 
When doing so at $Re_n=115$ (Fig.~\ref{f:PIV_N1}), small differences are observed for the aligned geometry, while the differences are more clearly detected for the opposed geometry. In any case, the flow does not show fully turbulent characteristics, although these results and pressure drop ones imply that the flow is, at least, transitional for both geometries.


The comparison of average and instantaneous velocity fields at $Re_n = 460$ (Fig.~\ref{f:PIV_N3}) confirms, for both geometries, the chaotic  nature of the flow. These observations can be quantified by performing a Proper Orthogonal Decomposition analysis \cite{sirovich} that decomposes the flow in a series of coherent structures (modes) arranged by its corresponding kinetic energy. The first mode corresponds to the averaged velocity field and, consequently, the kinetic energy of the averaged field can be compared to the total kinetic energy of the flow (averaged plus fluctuations) in order to evaluate the weight of the fluctuations.

\begin{figure}[htbp]
\centering
\includegraphics[width=8cm]{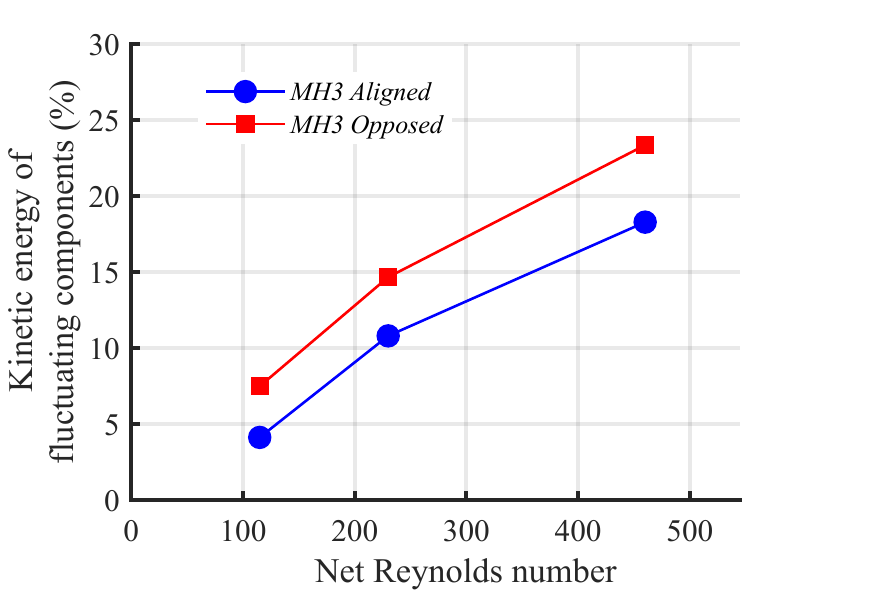}
\caption{Energy of the fluctuating components of the flow, from a Proper Orthogonal Decomposition analysis, as a function of the net Reynolds number}
\label{f:pod}
\end{figure}

Fig.~\ref{f:pod} represents the kinetic energy of the fluctuating components (as a percentage of the total kinetic energy of the flow) as a function of the net Reynolds number. Even at the lower net Reynolds number, $Re_n=110$, the fluctuating components have a significant weight on the energy of the flow for both configurations, $>5\%$. It should be considered that the flow may present fluctuations even under laminar flow conditions due to: measurement uncertainties, fluctuations of the flow rate due to the pump, etc. As a reference, a ~2\% of the kinetic energy of the fluctuations was measured for a laminar case in a one-orifice baffle configuration under similar working conditions (lighting, particle size, PIV algorithm, etc.) \cite{MunozPOD}.

The effect of increasing the net Reynolds numbers is clear for both geometries: an increase on the kinetic energy of the fluctuations. For the range of net Reynolds numbers tested, the weight of the fluctuations is higher for the opposed baffles (around 3-5\%), which confirms the more 'chaotic' behavior observed for this configuration.

Additionally, a better understanding of the instantaneous flow field characteristics for both geometries and the experiments N1 to N3 can be obtained from Video 1, where compositions of PIV instantaneous velocity fields are shown and analysed.


In contrast with the flow structures reported in one-orifice baffled tubes, low velocity regions do not appear downstream of each orifice in any of the configurations analysed. The recirculations close to the tube wall are not detected either, and this kind of structure is only found in the spaces opposed to each one of the orifices. This observation was already reproduced numerically by González-Juárez et al. \cite{Gonzalez} for oscillatory flow conditions.

\begin{figure}[htbp]
\subcaptionbox{\label{f:Nu_net}}[8cm]{\includegraphics[width=8cm]{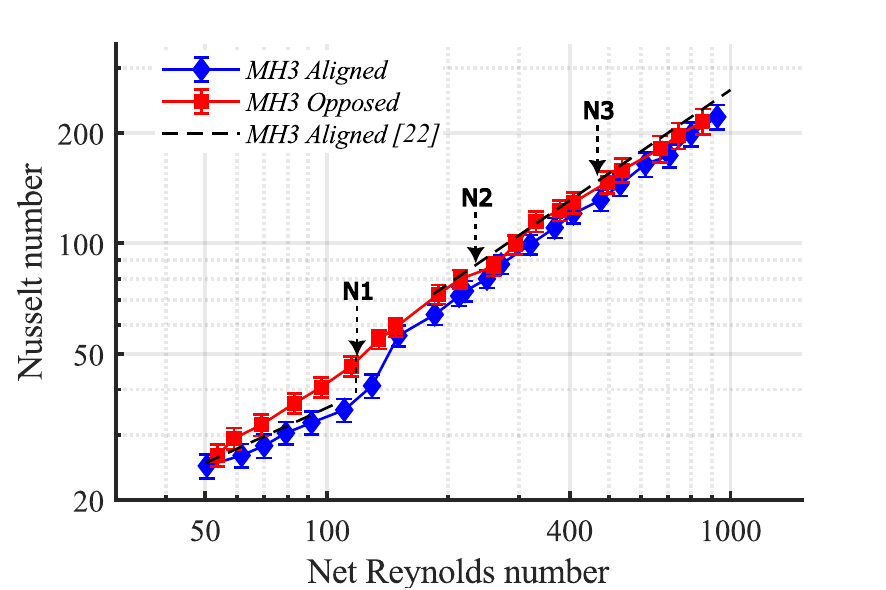}}\\
\subcaptionbox{\label{f:DeltaT_net}}[8cm]{ \includegraphics[width=8cm]{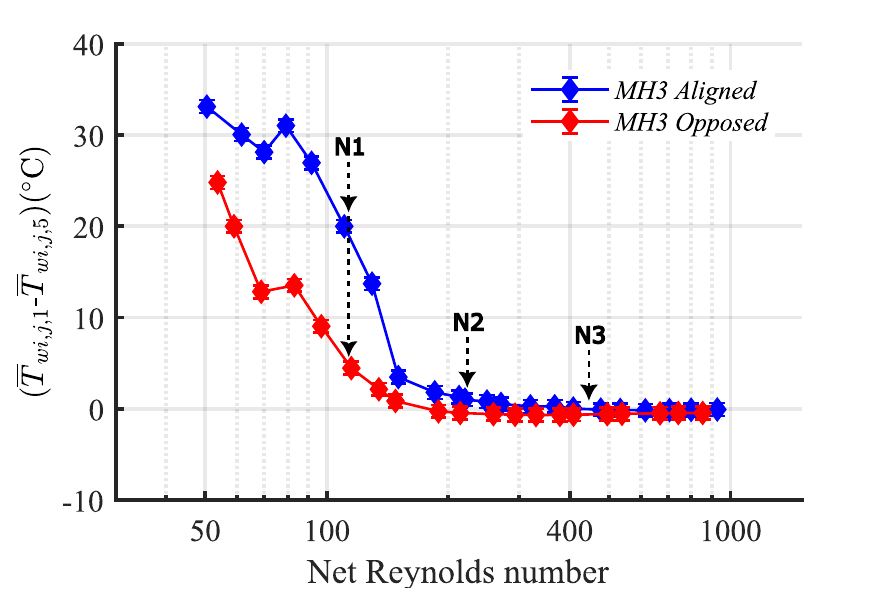}}
\caption{(a) Nusselt number and (b) temperature stratification vs net Reynolds number for both MH3 baffle orientations, $Pr=65$ (C=50\%). \label{f:Nu_DeltaT_net}}
\end{figure}

\paragraph{Heat transfer}
Fig.~\ref{f:Nu_net} shows the Nusselt number as a function of the net Reynolds number for both baffle orientations and a Prandtl number $Pr=65$. Only one propylene-glycol water mixture has been used, reducing the range tested in comparison to the pressure drop measurements. Rayleigh number in the range $Ra^*=1.5-3.2\cdot10^8$ has been tested; for the sake of a better comparison, the value is approximately the same for both baffle orientations at each net Reynolds number. 

As a reference, the dashed lines represent the values obtained from correlations for MH3 aligned baffles, available in a previous study  \cite{MunozMH3}. In the laminar flow regime, the experimental results overlap with the correlation, while the experimental data in the turbulent region falls below the correlation ($\sim$10\%). This mismatch is acceptable if we consider the uncertainty associated to the correlation, and the fact that it was developed for the range $190<Pr<470$, far from the value used in this study ($Pr=65$).

As can be seen, there is no variation in the curve slope for the range tested for the opposed baffles, confirming that no relevant flow transition takes place. In fact, all the data can be fitted to the correlation in Eq.~\ref{eq_nunet} with a deviation lower than 10\%. On the other hand, the aligned baffles show a sharp increase on the Nusselt number at $Re_n=130$, indicating the end of laminar flow regime. This value is close to the critical Reynolds number, $Re_n=105$, obtained by Muñoz-Cámara et al. \cite{MunozMH3}.

\begin{equation}
Nu_{op}=0.412\cdot Re_n^{0.757} \cdot Pr^{0.285}
\label{eq_nunet}
\end{equation}

It is the range from the end of the laminar region of the opposed baffles to the end of the laminar region of the aligned, $Re_n=40-130$, where there is a noticeable heat transfer enhancement. A maximum increase of 27\% in the Nusselt number is achieved at $Re_n=120$. For $Re_n>150$, the enhancement falls below the experimental uncertainty.

The different value for the critical net Reynolds number from the pressure drop results, $Re_n=90$, and the heat transfer results, $Re_n=130$, can be partly explained by the measurement uncertainty. But another relevant factor could be the thermal effects associated to the heat transfer tests: presence of mixed convection and wall-fluid temperature drop.

The existence of buoyancy effects on the flow can be deducted from local wall temperature measurements. As an example, the average temperature difference (for the 8 axial positions)  between the upper and lower thermocouples is plotted in Fig.~\ref{f:DeltaT_net} as a function of the net Reynolds number. 

Both baffle orientations display a high vertical temperature gradient at low net Reynolds numbers. The aligned baffles show the expected behaviour: temperature stratification (i.e., natural convection effects) in the laminar region ($Re_n<130$) and negligible stratification in the transitional-turbulent regions. 
On the contrary, the opposed baffles should not experience a flow transition in the tested range, but they also display a significantly high temperature stratification, but lower than the aligned baffles. Thus, it can be concluded that the opposed baffles are working under turbulent or transitional mixed convection conditions for $Re_n<130$. 

If the values for both configurations are compared, the opposed baffles display a significantly lower temperature stratification for all the range tested below $Re_n=130$, for example, for N1 ($Re_n=115$) the temperature difference is 15 $\degree$C lower for the opposed baffles. For higher net Reynolds numbers (for example points N2 and N3) the temperature stratification is negligible for both configurations.

There is a change in the decreasing trend of the temperature stratification as a function of the net Reynolds number, which cannot be explained by the measurement uncertainty. The increase in the temperature difference between $Re_n=70$ and $Re_n=80$ is due to the increase of the Rayleigh number from $Ra^*=1.6\cdot10^7$ to $Ra^*=2.2\cdot10^7$. While the Rayleigh number was kept constant for the three lowest net Reynolds numbers, it was modified for the next point to keep a high enough wall-fluid temperature to ensure an acceptable uncertainty for the Nusselt number.

\subsection{Oscillatory flow}


The oscillatory flow has been studied by performing pressure drop and visualization tests. 

\begin{figure}[htbp]
\centering
\includegraphics[width=8cm]{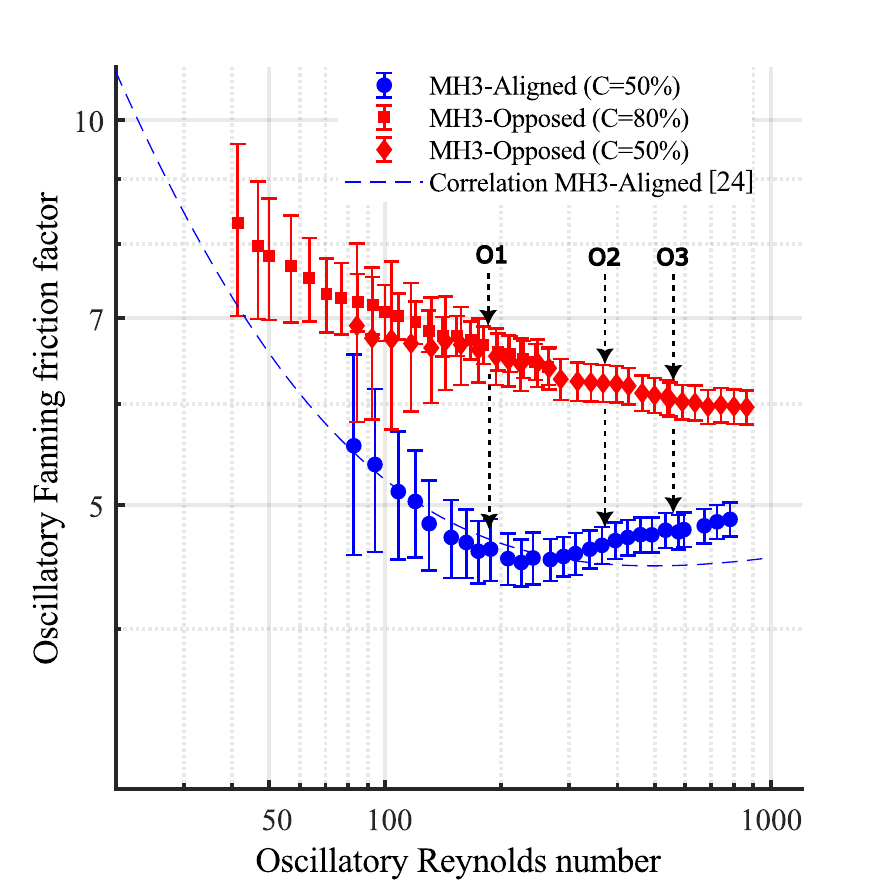}
\caption{Oscillatory Fanning friction factor vs oscillatory Reynolds number for both MH3 baffle orientations. $x_0/D=0.5$
\label{f:f_osc}
}
\end{figure}

\paragraph{Pressure drop}
Fig.~\ref{f:f_osc} shows the oscillatory Fanning friction factor as a function of the oscillatory Reynolds number for a dimensionless amplitude $x_0/D=0.5$. The oscillatory friction factor for the opposed baffles' configuration does not show any sharp change on its trend, indicating that there is no flow regime transition in the studied range, or it is quite smooth. For the aligned baffles' arrangement, friction reaches a minimum at around $Re_{osc}\approx200$ (near O1), which could correspond to the end of the unsteady laminar region. The friction factor is again significantly higher for the opposed baffles' configuration, around 20\% for $Re_{osc}=570$ (test 03). 

As a reference, results from the correlation proposed by Muñoz-Cámara et al. \cite{MunozPower}, which provides the oscillatory friction factor as a function of $Re_{osc}$ and $x_0/D$, is plotted. For $Re_{osc}<300$, there is a good match between the experimental results and the correlation. However, the correlation underestimates the friction factor at high $Re_{osc}$, what could be due to: (1) the correlation was developed for $0.14<x_0/D<0.43$ while the value $x_0/D=0.5$ was used for this study, and (2) the relative lack of points at high $Re_{osc}$ that were available to develop the correlation. Regarding the opposed baffles, the data seems to approach the aligned baffles curve at low $Re_{osc}$, as it happens for net flow conditions, however the minimum $Re_{osc}$ tested is not low enough to confirm this point.

The data for the opposed MH3 baffles can be fitted to the following correlation within a 5\% deviation:

\begin{equation}
f_{osc,op}=\frac{70.1}{Re_{osc}}+7.32 \cdot Re_{osc}^{-0.0327}, \\\\x_0/D=0.53
\label{eq_fosc_op}
\end{equation}

The uncertainty associated to the experimental measurements is high, and it is mainly due to the pressure transducer uncertainty. However, the real uncertainty is expected to be lower due to the fact that the differential pressure transducer measures the amplitude of the pressure drop (i.e., the difference between the pressure drop during the positive and negative half-cycles), instead of an 'absolute' pressure; this implies that the bias component of the sensor error would have a lower effect on the measurement error than the one that was considered.  

\begin{figure}[htbp]
\centering
\includegraphics[width=7 cm]{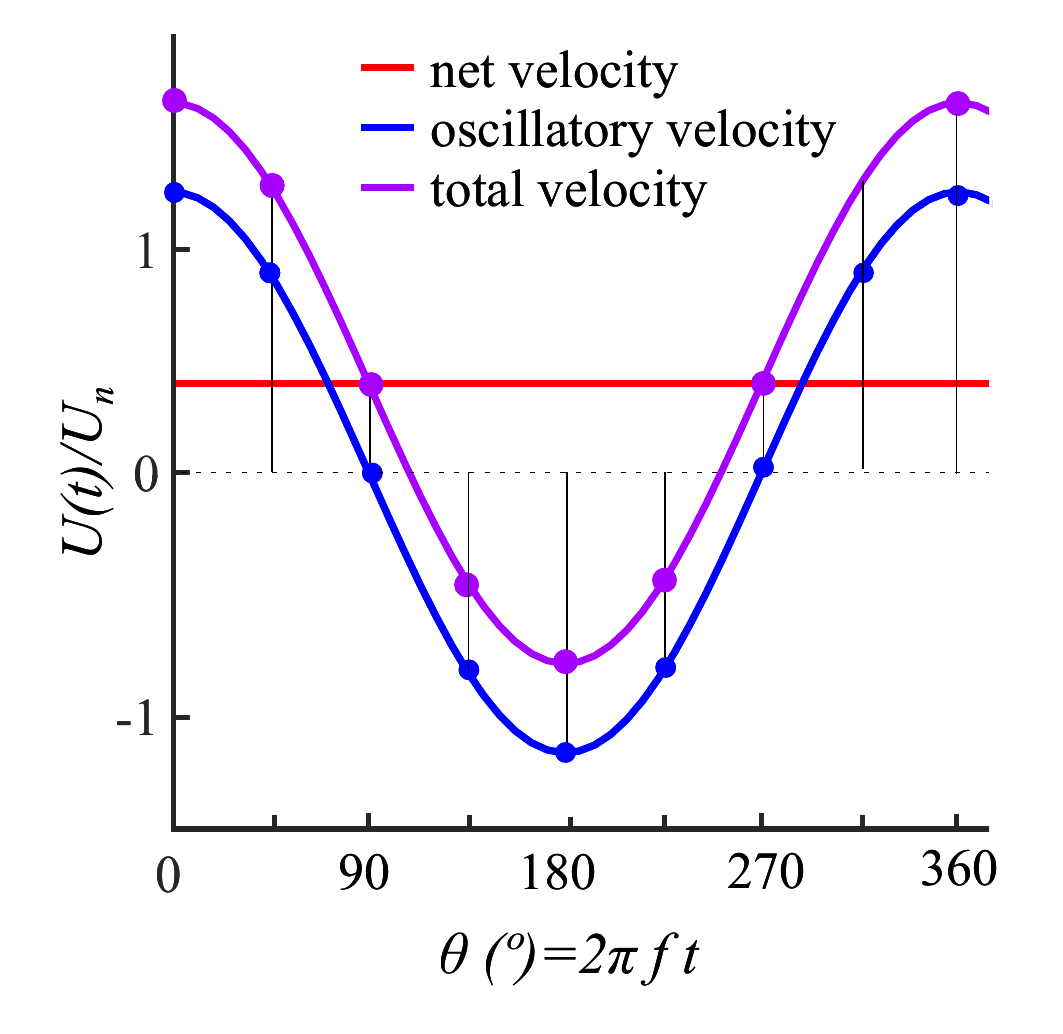}
\caption{ Phases of the oscillation cycle for flow field representation (PIV technique).\label{f:PIV_phases}}
\end{figure}
\begin{figure*}[htbp]
\centering
\subcaptionbox{Aligned baffle arrangement.\label{f:PIV_O1_ALI}}[\textwidth]{ 
\begin{overpic}[width=\anchoPIV]{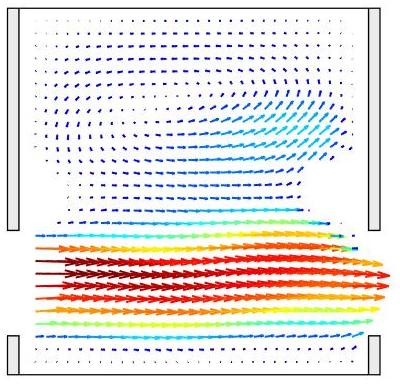} 
\put (\labelpPIVAB,\labelPIVver)
{\labelPIVA}



\end{overpic} 
\begin{overpic}[width=\anchoPIV]{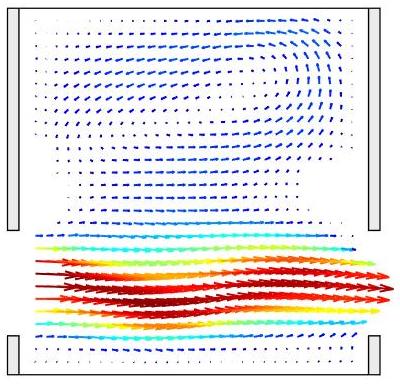} 
\put (\labelpPIVAB,\labelPIVver) {\labelPIVB}
\end{overpic} 
\begin{overpic}[width=\anchoPIV]{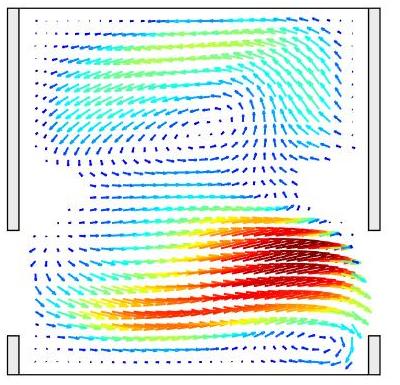}
\put (\labelpPIVCD,\labelPIVver) {\labelPIVC}
\put(55,-5){\RegFlujo \vector(0,1){20}~ I}

\end{overpic} 
\begin{overpic}[width=\anchoPIV]{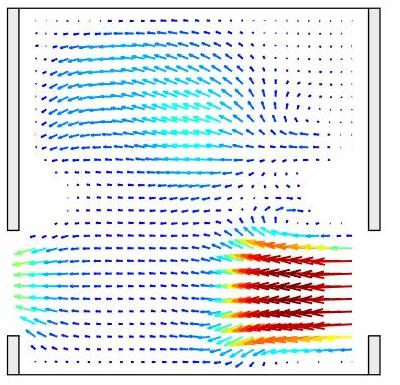}
\put (\labelpPIVCD,\labelPIVver) {\labelPIVD}
\put(70,-5){\RegFlujo \vector(0,1){15}~ II}
\end{overpic}}\\

\espacioFigPIV

\subcaptionbox{Opposed baffle arrangement. \label{f:PIV_O1_DES}}[\textwidth]{ 
\begin{overpic}[width=\anchoPIV]{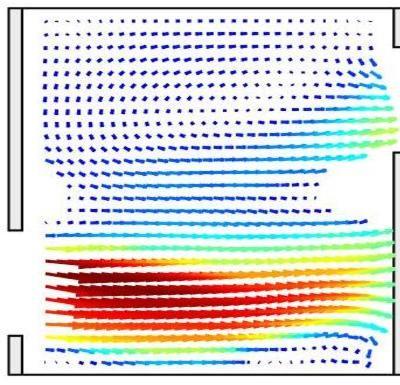} 
\put (\labelpPIVAB,\labelPIVver) {\labelPIVA}

\end{overpic} 
\begin{overpic}[width=\anchoPIV]{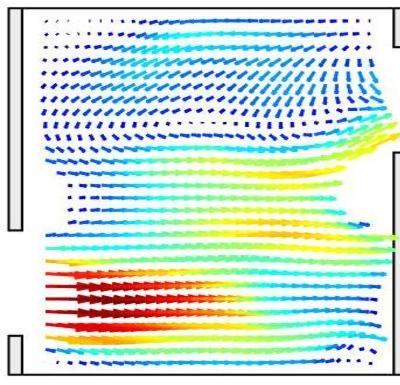} 
\put (\labelpPIVAB,\labelPIVver) {\labelPIVB}
\end{overpic} 
\begin{overpic}[width=\anchoPIV]{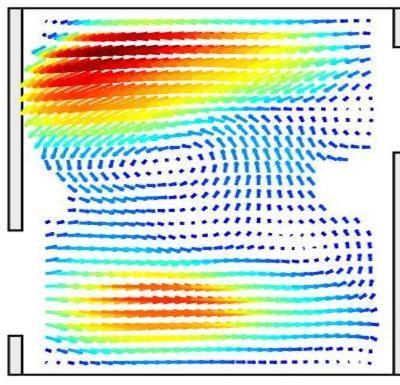}
\put (\labelpPIVCD,\labelPIVver) {\labelPIVC}
\end{overpic} 
\begin{overpic}[width=\anchoPIV]{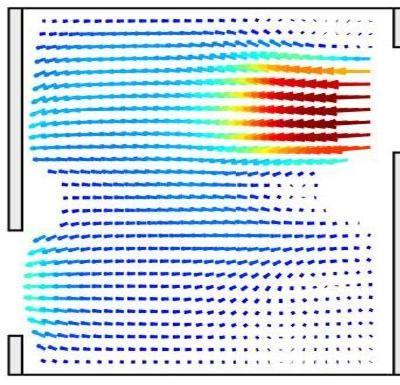}
\put (\labelpPIVCD,\labelPIVver) {\labelPIVD}
\end{overpic}}\\
\barracolor
\caption{Phase-average pure oscillatory flow field at $Re_{osc}=190$ for both geometries and different cycle positions. Experiment O1 (Table~\ref{table_piv}). \label{f:PIV_O1}}
\end{figure*}
\begin{figure*}[htbp]
\centering
\subcaptionbox{Aligned baffle arrangement.\label{f:PIV_O3_ALI}}[\textwidth]{ 
\begin{overpic}[width=\anchoPIV]{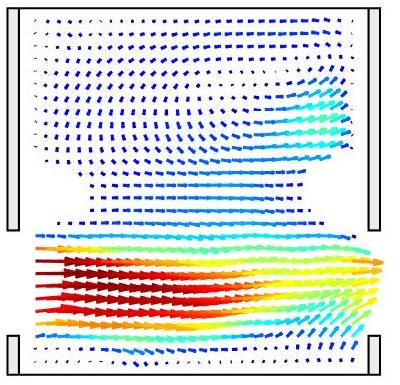} 
\put (\labelpPIVAB,\labelPIVver) {\labelPIVA}


\end{overpic} 
\begin{overpic}[width=\anchoPIV]{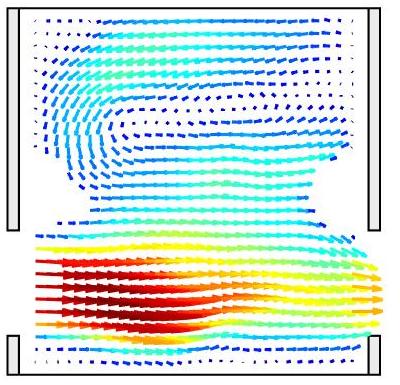} 
\put (\labelpPIVAB,\labelPIVver) {\labelPIVB}
\end{overpic} 
\begin{overpic}[width=\anchoPIV]{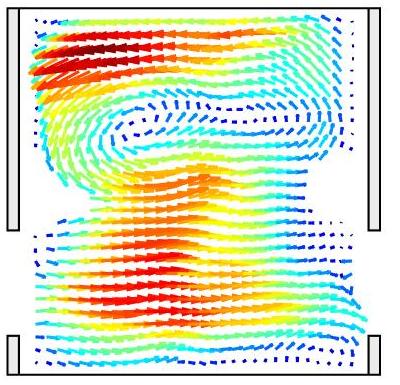}
\put (\labelpPIVCD,\labelPIVver) {\labelPIVC}
\end{overpic} 
\begin{overpic}[width=\anchoPIV]{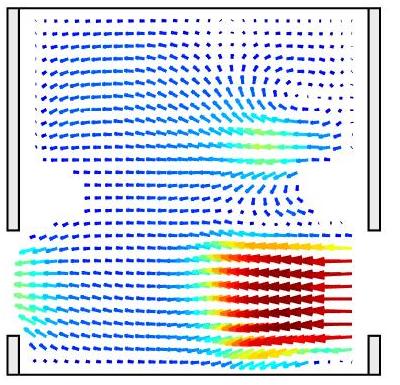}
\put (\labelpPIVCD,\labelPIVver) {\labelPIVD}
\end{overpic}}\\

\espacioFigPIV

\subcaptionbox{Opposed baffle arrangement. \label{f:PIV_O3_DES}}[\textwidth]{ 
\begin{overpic}[width=\anchoPIV]{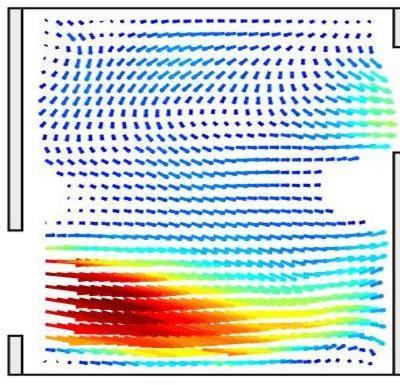} 
\put (\labelpPIVAB,\labelPIVver) {\labelPIVA}
\end{overpic} 
\begin{overpic}[width=\anchoPIV]{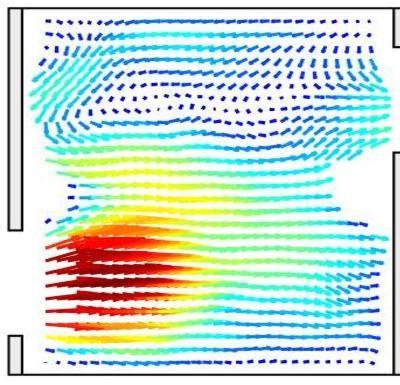} 
\put (\labelpPIVAB,\labelPIVver) {\labelPIVB}
\end{overpic} 
\begin{overpic}[width=\anchoPIV]{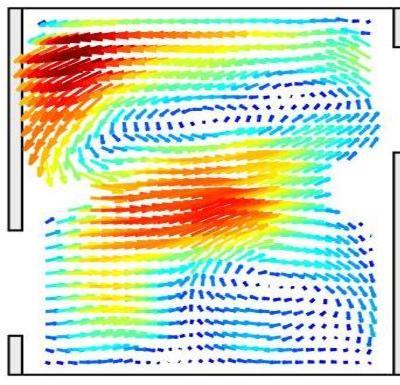}
\put (\labelpPIVCD,\labelPIVver) {\labelPIVC}
\end{overpic} 
\begin{overpic}[width=\anchoPIV]{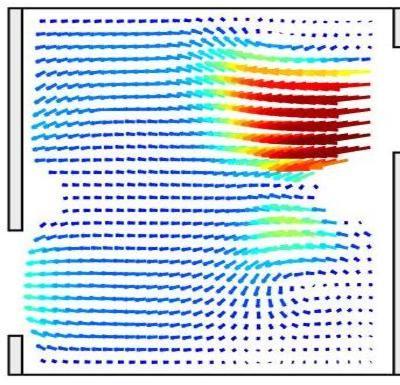}
\put (\labelpPIVCD,\labelPIVver) {\labelPIVD}
\end{overpic}}\\
\barracolor
\caption{Phase-average pure oscillatory flow field at $Re_{osc}=570$ for both geometries and different cycle positions. Experiment O3 (Table~\ref{table_piv}).\label{f:PIV_O3}}
\end{figure*}

\paragraph{Velocity field}
The oscillatory flow field is presented for both configurations and $Re_{osc}= 190$ (O1) and $Re_{osc}= 570$ (O3) in Fig.~\ref{f:PIV_O1} and Fig.~\ref{f:PIV_O3}, respectively.
The flow pattern of the experiments is shown for different positions of the oscillatory cycle, which is presented in terms of $\theta$, the cycle position, as illustrated in Fig.~\ref{f:PIV_phases}. The value $\theta=0\degree$ corresponds to the instant of maximum average velocity towards the right side.

The oscillatory flow is symmetric in time, meaning that the flow between $0\degree<\theta<180\degree$ is symmetric to the one between  $180\degree<\theta<360\degree$. Being so, results are presented for $\theta=0\degree,~45\degree,~90\degree,~135\degree$. 
The results for $Re_{osc}= 190$ at $\theta=0\degree$ (maximum averaged velocity) in the aligned geometry (Fig.~\ref{f:PIV_O1_ALI}), show a phase-average velocity field which is very similar to the one of the net flow: a jet connects consecutive baffles orifices and a recirculation is present in the opposed flow region. At $\theta=90\degree$, when the average velocity equals zero, the flow pattern is still governed by the jet due to the inertia of the flow (Fig.~\ref{f:PIV_O1_ALI}-I). Later in the cycle, at $\theta=135\degree$, the previous jet has vanished, and a new jet is formed at the right baffle (Fig.~\ref{f:PIV_O1_ALI}-II). 

For the opposed baffle geometry, the results show similar trends (Fig.~\ref{f:PIV_O1_DES}), beginning at $\theta=0\degree$ with a pattern similar to the one obtained for the net flow test, with a jet in the direction of the flow which loses strength and at $\theta=90\degree$ is part of a recirculation which fully covers the inter-baffle space. At $\theta=135\degree$, a new jet takes shape in the opposite direction, from the baffle orifice at the right-hand side of the picture. 

The videos (for points O1, O2 and O3 and both configuration)  support the trend of the pressure drop results: the friction factor is deviated from the laminar trend for $Re_{osc}>200$ and, at the same time, the instantaneous velocity fields along one oscillation cycle display a more chaotic behaviour with the jet deviating from its direction and fluctuating.


PIV phase-average velocity fields for  $Re_{osc}=570$ are shown in Fig.~\ref{f:PIV_O3}. When compared with results for $Re_{osc}=190$, the increase in the oscillatory Reynolds number  implies an increase of momentum diffusivity. Other than that, the non-dimensional average velocity fields are similar for both working conditions.

The oscillatory Reynolds number tested, $Re_{osc}= 570$, is very close to that studied by Nogueira et al. \cite{Nogueira} for aligned baffles. While the amplitude they tested was low, $x_0/D=0.045$, in comparison to the value used in this study, $x_0/D=0.5$, they also observed a three-dimensional and very complex flow for $Re_{osc}= 600$, which was dominated by eddy formation, as can be also observed in the video for the test O3 (aligned configuration).

\subsection{Compound flow}

In this section, the superposition of net and oscillatory flows is studied using visualization and heat transfer experiments. In order to quantify their relative importance, the velocity ratio is defined as follows:

\begin{equation}
    \Psi=\frac{Re_{osc}}{Re_n}
\end{equation}


\paragraph{Velocity field}
PIV experiments have been carried out for a fixed net Reynolds number of $Re_n=115$ and an oscillatory Reynolds number which varies within the range $Re_{osc}=190-570$, resulting in a velocity ratio within $\Psi=1.7-5.0$. 

\begin{figure*}[htbp]
\centering
\netflowdir
\subcaptionbox{Aligned baffle arrangement.\label{f:PIV_C1_ALI}}[\textwidth]{ 
\begin{overpic}[width=\anchoPIV]{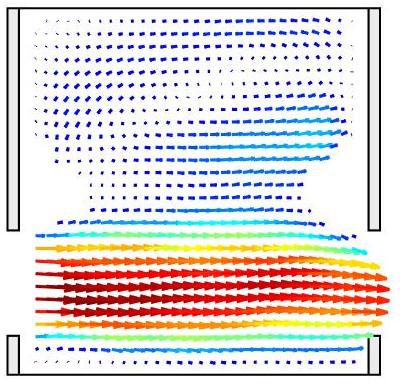} 
\put (\labelpPIVAB,\labelPIVver) {\labelPIVA}
\end{overpic} 
\begin{overpic}[width=\anchoPIV]{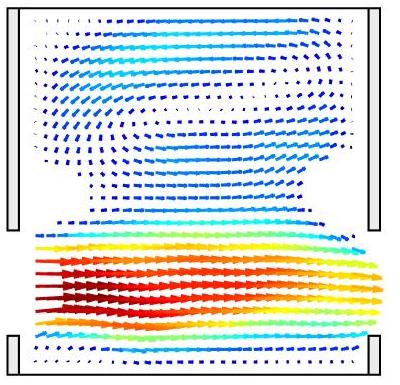} 
\put (\labelpPIVAB,\labelPIVver) {\labelPIVB}
\end{overpic} 
\begin{overpic}[width=\anchoPIV]{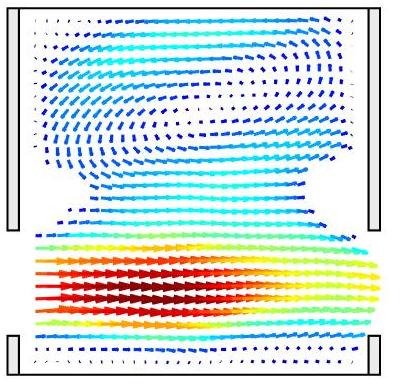}
\put (\labelpPIVCD,\labelPIVver) {\labelPIVC}
\end{overpic} 
\begin{overpic}[width=\anchoPIV]{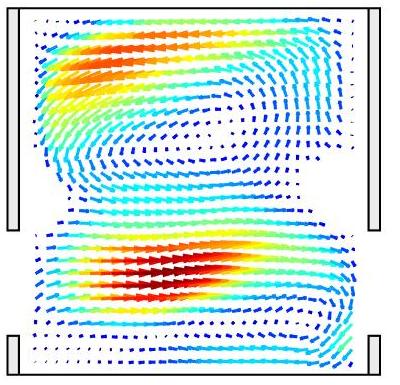}
\put (\labelpPIVCD,\labelPIVver) {\labelPIVD}
\end{overpic} \\

\espacioFigPIV

\begin{overpic}[width=\anchoPIV]{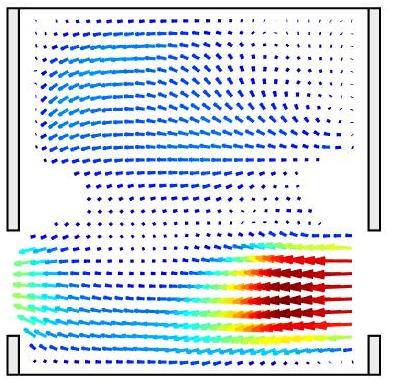} 
\put (\labelpPIVAB,\labelPIVver) {\labelPIVE}
\end{overpic} 
\begin{overpic}[width=\anchoPIV]{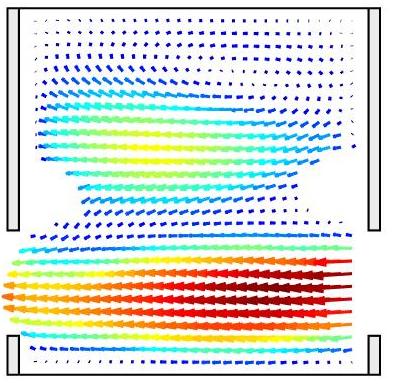} 
\put (\labelpPIVAB,\labelPIVver) {\labelPIVF}
\put(30,-5){\RegFlujo I~\vector(0,1){15}}
\put(28,-3){\RegFlujo \vector(-2,1){43}}
\end{overpic} 
\begin{overpic}[width=\anchoPIV]{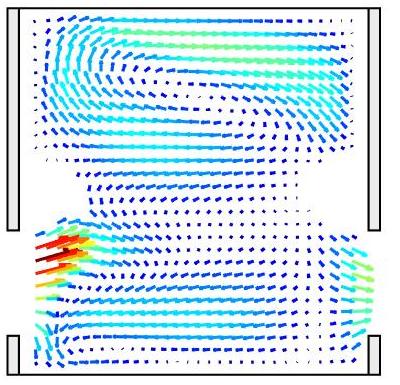}
\put (\labelpPIVCD,\labelPIVver) {\labelPIVG}
\end{overpic} 
\begin{overpic}[width=\anchoPIV]{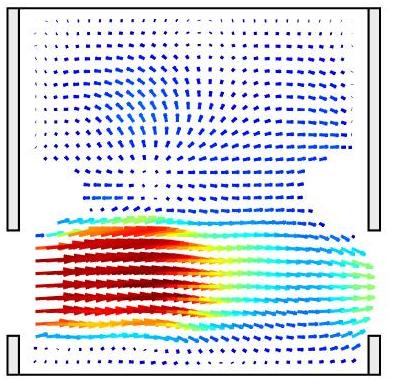}
\put (\labelpPIVCD,\labelPIVver) {\labelPIVH}
\end{overpic}
}\\

\espacioFigPIV

\subcaptionbox{Opposed baffle arrangement. \label{f:PIV_C1_DES}}[\textwidth]{ 
\begin{overpic}[width=\anchoPIV]{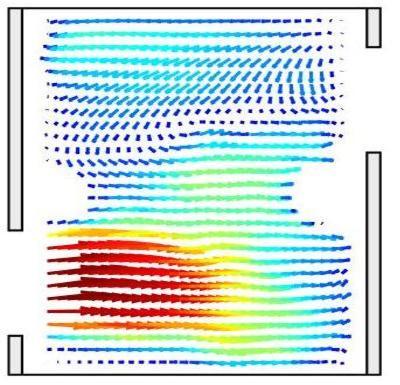} 
\put (\labelpPIVAB,\labelPIVver) {\labelPIVA}
\end{overpic} 
\begin{overpic}[width=\anchoPIV]{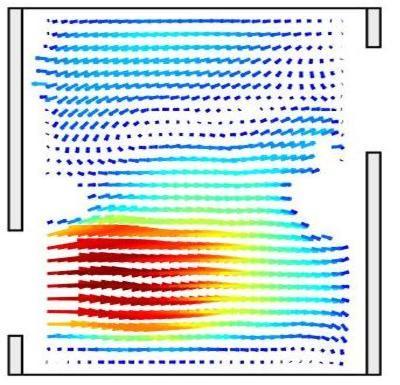} 
\put (\labelpPIVAB,\labelPIVver) {\labelPIVB}
\end{overpic} 
\begin{overpic}[width=\anchoPIV]{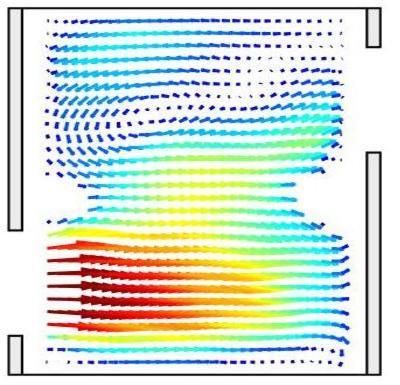}
\put (\labelpPIVCD,\labelPIVver) {\labelPIVC}
\end{overpic} 
\begin{overpic}[width=\anchoPIV]{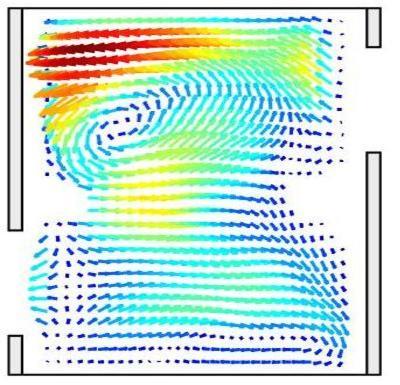}
\put (\labelpPIVCD,\labelPIVver) {\labelPIVD}
\end{overpic}\\

\espacioFigPIV

\begin{overpic}[width=\anchoPIV]{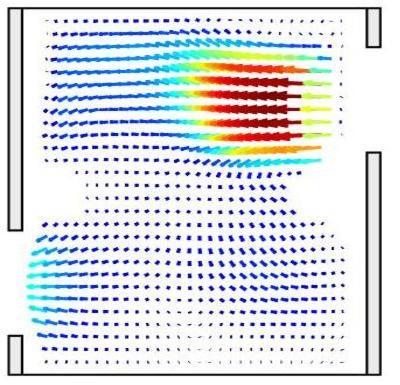} 
\put (\labelpPIVAB,\labelPIVver) {\labelPIVE}
\end{overpic} 
\begin{overpic}[width=\anchoPIV]{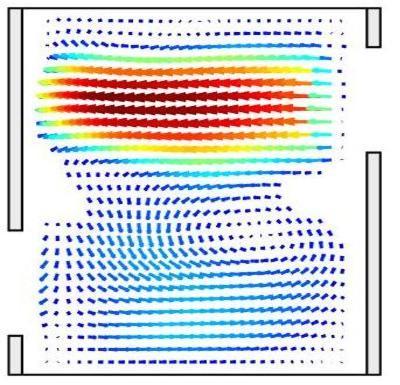} 
\put (\labelpPIVAB,\labelPIVver) {\labelPIVF}
\end{overpic} 
\begin{overpic}[width=\anchoPIV]{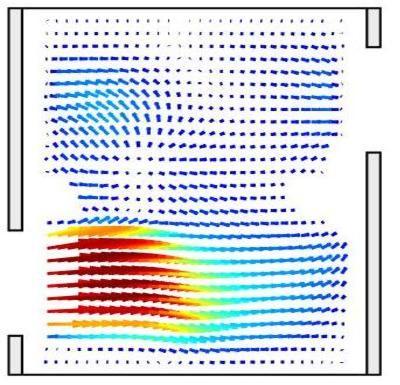}
\put (\labelpPIVCD,\labelPIVver) {\labelPIVG}
\end{overpic} 
\begin{overpic}[width=\anchoPIV]{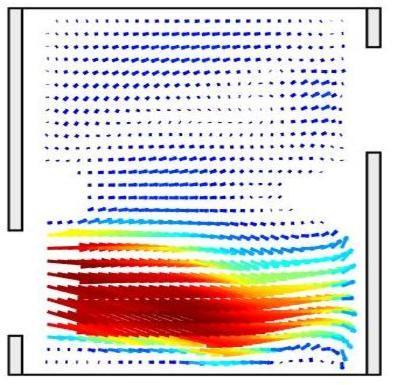}
\put (\labelpPIVCD,\labelPIVver) {\labelPIVH}
\end{overpic}}\\
\barracolor
\caption{Phase-average compound flow field at $Re_n=115$ (flowing rightwards) and $\Psi = 1.7$ for both geometries and different cycle positions. Experiment C1 (Table~\ref{table_piv}). \label{PIV_C1}}
\end{figure*}

Fig.~\ref{PIV_C1} shows the phase-averaged velocity field for both aligned and opposed baffles arrangements at $\Psi = 1.7$ (C1). In this case, the net flow bulk velocity is lower than the maximum oscillatory flow bulk velocity. This results in a change of direction of the compound flow. In the figure, the net flow goes rightwards. The result for both geometries is that the flow is still dominated by the  net flow direction, with flow reversal (Fig.~\ref{f:PIV_C1_ALI}-I) taking place for a small fraction of the cycle $\theta=180\degree-225\degree$.

As before, with regard to convection mechanisms, we observe recirculations at almost any cycle position, improving the radial mixing. Flow circulation in radial direction is relatively more important at positions where a change of flow direction occurs (at $\theta=135\degree,~270\degree$), proving that flow reversal is critical to achieve a good mixing when there is a net flow. 


\begin{figure*}[htbp]
\centering
\netflowdir
\subcaptionbox{Aligned baffle arrangement.\label{f:PIV_C3_ALI}}[\textwidth]{ 
\begin{overpic}[width=\anchoPIV]{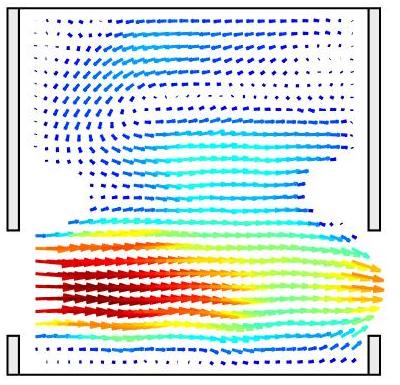} 
\put (\labelpPIVAB,\labelPIVver) {\labelPIVA}
\end{overpic} 
\begin{overpic}[width=\anchoPIV]{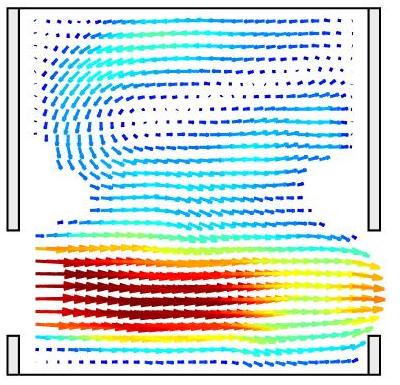} 
\put (\labelpPIVAB,\labelPIVver) {\labelPIVB}
\end{overpic} 
\begin{overpic}[width=\anchoPIV]{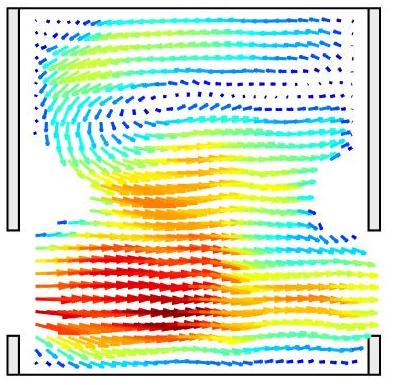}
\put (\labelpPIVCD,\labelPIVver) {\labelPIVC}
\end{overpic} 
\begin{overpic}[width=\anchoPIV]{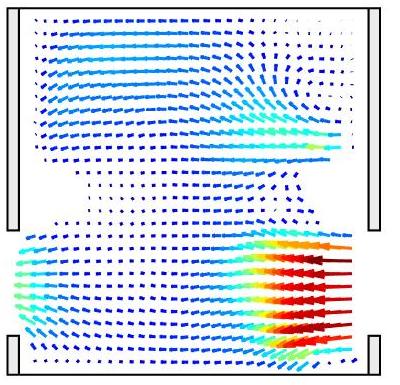}
\put (\labelpPIVCD,\labelPIVver) {\labelPIVD}
\end{overpic} \\

\espacioFigPIV

\begin{overpic}[width=\anchoPIV]{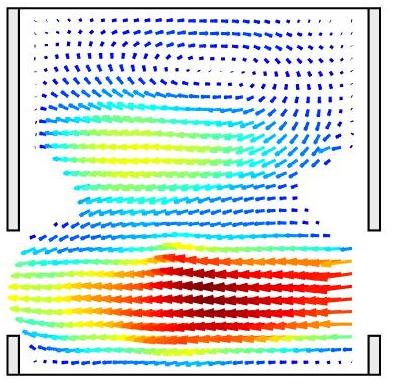} 
\put (\labelpPIVAB,\labelPIVver) {\labelPIVE}
\end{overpic} 
\begin{overpic}[width=\anchoPIV]{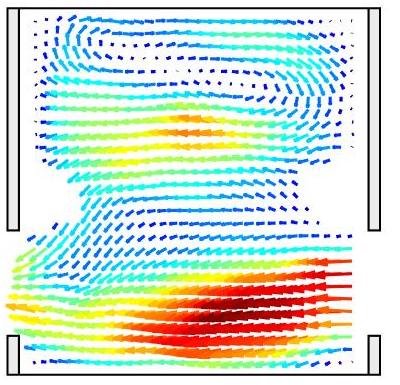} 
\put (\labelpPIVAB,\labelPIVver) {\labelPIVF}
\end{overpic} 
\begin{overpic}[width=\anchoPIV]{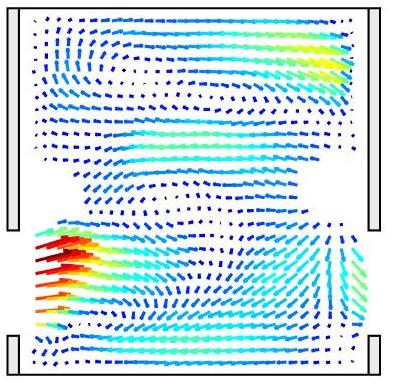}
\put (\labelpPIVCD,\labelPIVver) {\labelPIVG}
\end{overpic} 
\begin{overpic}[width=\anchoPIV]{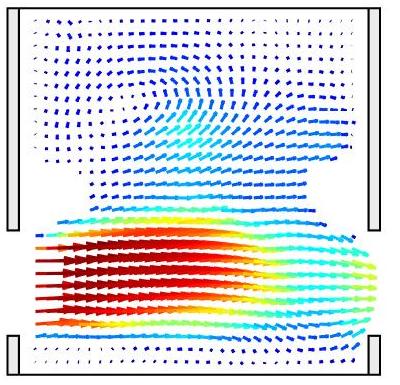}
\put (\labelpPIVCD,\labelPIVver) {\labelPIVH}
\end{overpic}
}\\

\espacioFigPIV

\subcaptionbox{Opposed baffle arrangement. \label{f:PIV_C3_DES}}[\textwidth]{ 
\begin{overpic}[width=\anchoPIV]{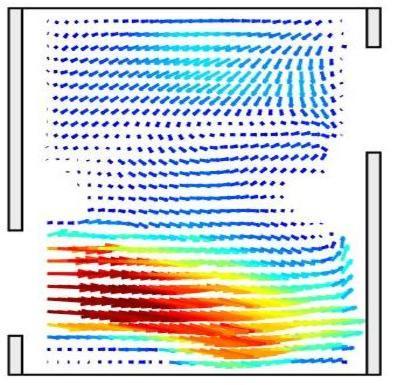} 
\put (\labelpPIVAB,\labelPIVver) {\labelPIVA}
\end{overpic} 
\begin{overpic}[width=\anchoPIV]{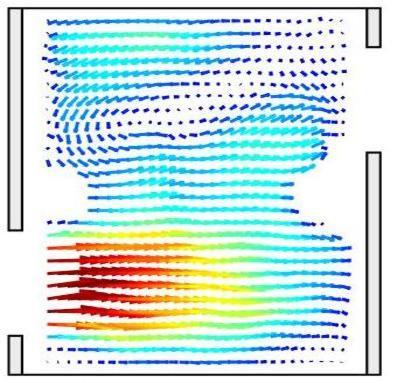} 
\put (\labelpPIVAB,\labelPIVver) {\labelPIVB}
\end{overpic} 
\begin{overpic}[width=\anchoPIV]{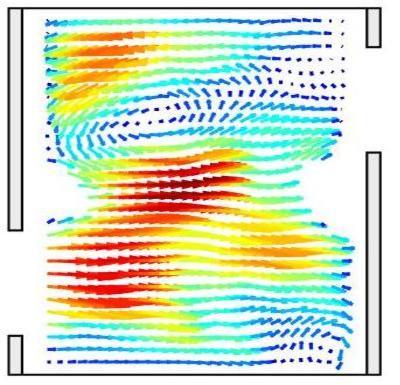}
\put (\labelpPIVCD,\labelPIVver) {\labelPIVC}
\end{overpic} 
\begin{overpic}[width=\anchoPIV]{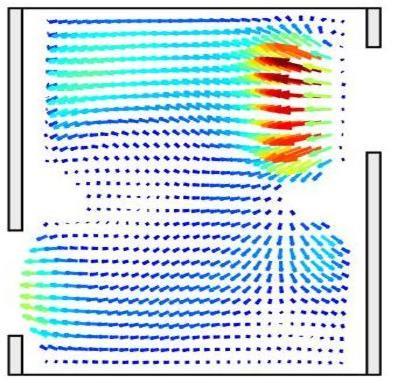}
\put (\labelpPIVCD,\labelPIVver) {\labelPIVD}
\end{overpic}\\

\espacioFigPIV

\begin{overpic}[width=\anchoPIV]{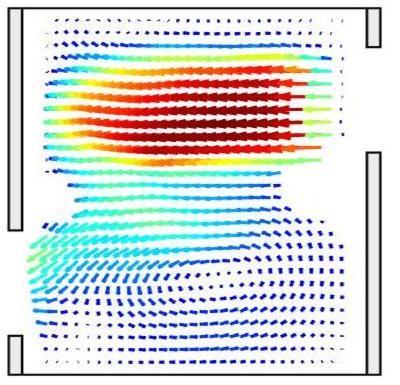} 
\put (\labelpPIVAB,\labelPIVver) {\labelPIVE}
\end{overpic} 
\begin{overpic}[width=\anchoPIV]{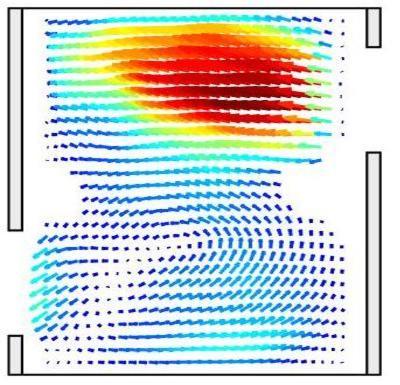} 
\put (\labelpPIVAB,\labelPIVver) {\labelPIVF}
\end{overpic} 
\begin{overpic}[width=\anchoPIV]{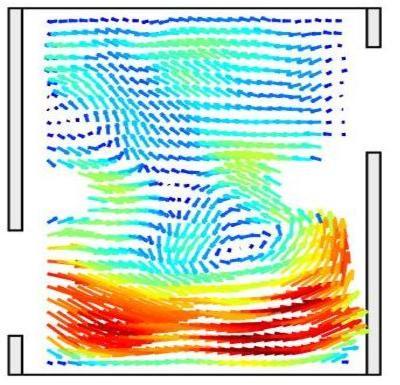}
\put (\labelpPIVCD,\labelPIVver) {\labelPIVG}
\end{overpic} 
\begin{overpic}[width=\anchoPIV]{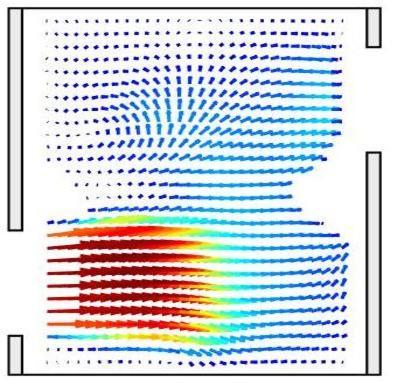}
\put (\labelpPIVCD,\labelPIVver) {\labelPIVH}
\end{overpic}}\\
\barracolor
\caption{Phase-average compound flow field at  $Re_n = 115$ (flowing rightwards),
 $\Psi = 5$ for both geometries and different cycle positions. Experiment C3 (Table~\ref{table_piv}). \label{PIV_C3}}
\end{figure*}

The increase of the relative importance of the oscillatory flow up to a velocity ratio $\Psi=5$ (C3) results in an increase of the fraction of the cycle with a reverse flow. Thus, the importance of the net flow is significantly reduced, and the results presented in Fig.~\ref{PIV_C3} show very similar flow patterns to the ones of a pure oscillatory flow (Fig.~\ref{f:PIV_O3}). However, the effect of the net flow is still noticeable, for example, at $\theta=90\degree$ the jet formed in the orifice is stronger than that of the oscillatory flow, or at $\theta=135\degree$, the jet in opposite direction is weaker in comparison to the oscillatory flow case.


The only study available in the open literature for aligned three-orifice baffles under compound flow conditions was performed by González-Juárez et al. \cite{Gonzalez}, where they simulated a flow with $Re_n=50$ and $Re_{osc}=800$, for an amplitude $x_0/D=0.3$. The authors observed a highly chaotic flow and vigorous mixing, which is consistent with the measured velocity fields at $Re_n=115$ and $Re_{osc}=570$ (see video of the test C3, aligned configuration), when the flow already displays similar characteristics.

\begin{figure}[htbp]
\centering
\subcaptionbox{$Re_{osc}\approx 150,~560$}[7cm]{
\includegraphics[width=8cm]{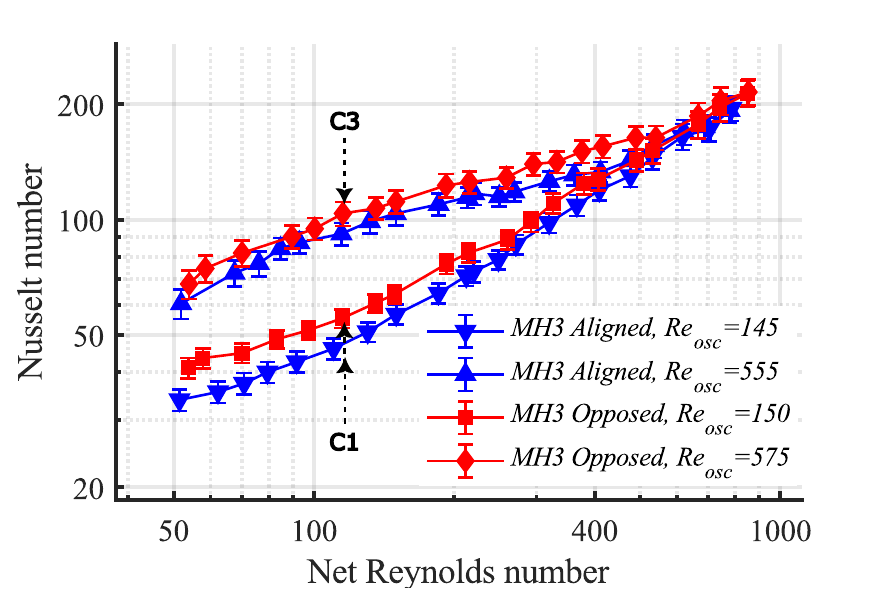}
}
\subcaptionbox{$Re_{osc}\approx 380,~750$}[7cm]{
\includegraphics[width=8cm]{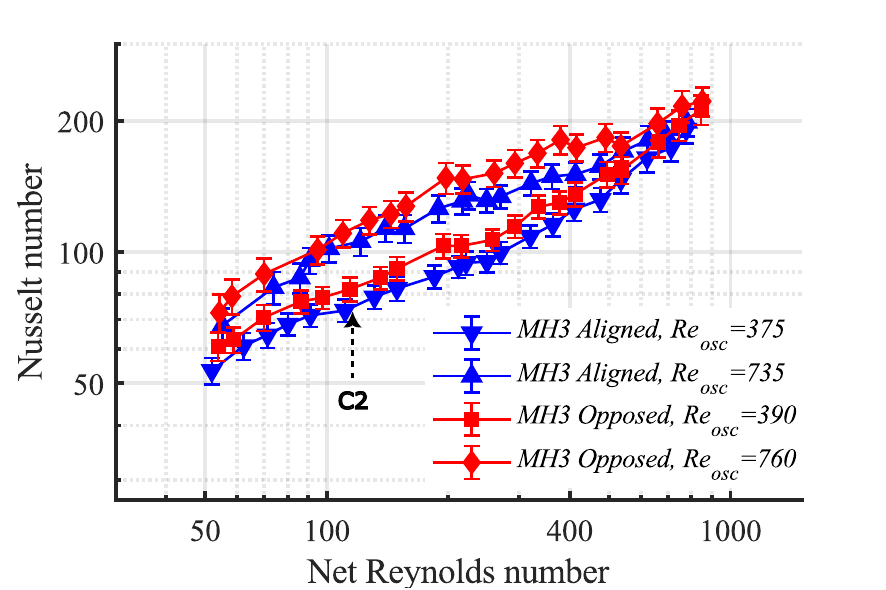}
}
\caption{Nusselt number vs net Reynolds number for different oscillatory Reynolds numbers and both MH3 baffle orientations. 
\label{f:Nu_osc1}
}
\end{figure}

\paragraph{Heat transfer}
Fig.\ref{f:Nu_osc1} a-b show the Nusselt number as a function of the net Reynolds number for both baffle orientations, and different oscillatory Reynolds numbers. The dimensionless amplitude is identical to that used for the pure oscillatory flow tests ($x_0/D=0.5$). For the sake of comparison, the Prandtl number, the Rayleigh number and the oscillatory Reynolds number for each test are kept almost identical for both baffles orientations.

The effect of the net Reynolds number on the Nusselt is noticeable even at low net Reynolds numbers and high oscillatory Reynolds numbers (high velocity ratios). This implies that the net component has an important role on the heat transfer process and cannot be neglected even for the highest velocity ratios tested ($\Psi\approx15$). This is supported by the flow patterns previously observed (Fig.~\ref{PIV_C1} and~\ref{PIV_C3}), where the net flow show a strong influence on the flow for low and high velocity ratios.

The oscillatory Reynolds number has also a positive effect on the heat transfer enhancement, e.g., a maximun increase of 100\% on the Nusselt number is observed at $Re_n=50$ when the oscillatory Reynolds number is increased from 150 to 760. The trend can also be clearly observed if we follow the values of the PIV tests: for C1 ($Re_n=115$, $Re_{osc}=190$) $Nu \approx 50$, while for C3 ($Re_n=115$, $Re_{osc}=570$) $Nu \approx 100$. However, all the curves overlap when the magnitude of the net and oscillatory Reynolds numbers is similar.

These trends can be presented more clearly if the data for the opposed baffles is fitted to a correlation with the following form:

\begin{equation}
Nu_{op}=0.385\cdot Re_n^{0.338} \cdot Re_{osc}^{0.435} \cdot Pr^{0.285}, \Psi>1
\label{eq_nuosc}
\end{equation}

where the data with $\Psi<1$ has not been considered because they follow the trend of the net flow tests.

In general, if baffle orientations are compared, the trends are almost identical for the four frequencies tested. In addition, while the opposed baffles provide consistently higher Nusselt number for all the range tested, the heat transfer enhancement is moderate and falls below the measurement uncertainty. Only for the lowest frequency tested ($Re_{osc}=145-150$), and low net Reynolds numbers ($Re_n<150$) there is a noticeable heat transfer enhancement (around 20\% at $Re_n=55$, but this value falls to a 10\% at $Re_n=250$).

\begin{figure}[htbp]
\centering
\includegraphics[width=8cm]{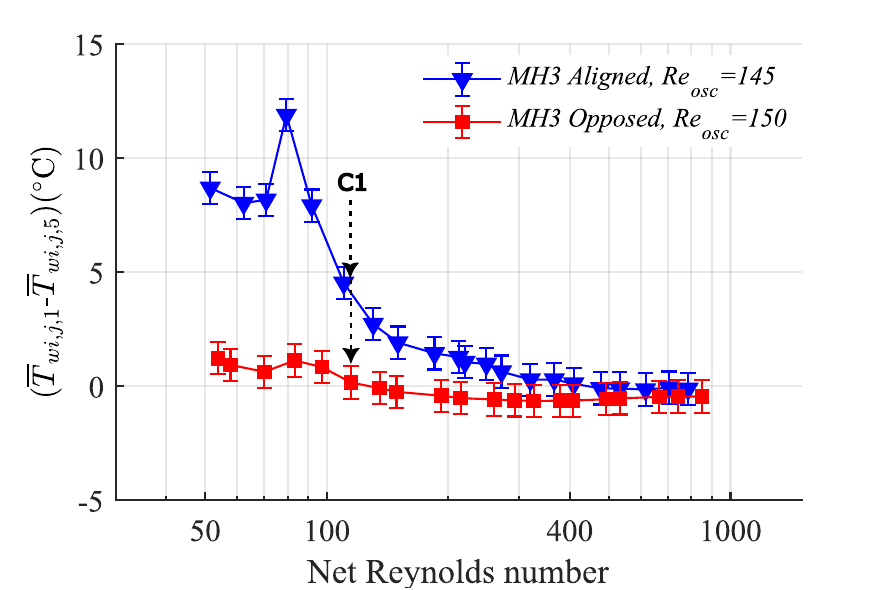}
\caption{Temperature stratification vs net Reynolds number for both MH3 baffle orientations. 
\label{f:estra_osc}
}
\end{figure}

The effect of the oscillatory flow can also be observed on the temperature stratification, as shown in Fig.~\ref{f:estra_osc}, where the upper-lower local wall temperature difference is plotted for both orientations at the lowest oscillatory Reynolds number tested, $Re_{osc}\approx 150$. For the aligned baffles, the temperature difference at low $Re_n$ is reduced up to a 70\%, from 30$^\circ$C to 8 $^\circ$C at $Re_n=60$, while for the opposed baffles the stratification is almost negligible for all the $Re_n$ range tested, supporting what was observed for net flow tests: the opposed baffles reduce significantly the temperature stratification when compared to the aligned. For both orientations, at higher oscillatory Reynolds numbers, the stratification is non-existent.

\section{Conclusions}

A systematic experimental analysis in a three-orifice oscillatory baffled tube has allowed to establish the effect of opposing consecutive baffles in their friction, heat transfer and flow pattern characteristics. The main conclusions derived from this work are:

\begin{itemize}

\item For net flow: opposed baffles increase the friction factor up to 40 \% in the range $40<Re_n<1000$, while heat transfer is only increased up to 27 \% in the range $50 < Re_n < 150$. This augmentation is caused by the earlier transition to turbulence reduction of the critical Reynolds number when the baffles are opposed, from $Ren \approx 100$ to $Ren \approx 40$. The visualization tests shows a more chaotic behaviour for the opposed baffles at low net Reynolds numbers.

\item For oscillatory flow: opposed baffles display a maximum increase of the oscillatory friction factor of 20 \% in comparison to the aligned baffles. Apparently, the deviation from the expected laminar trend for the opposed baffles takes place at a lower oscillatory Reynolds number. More tests for $Re_{osc}<50$ should be performed to prove this point. 

\item For compound flow, both baffle orientations show a remarkable effect of the oscillatory flow on the Nusselt number, up to a 100\% increase at the lower net Reynolds number tested, $Re_n=50$. However, there are only noticeable differences between both orientations at low Reynolds numbers: $Re_{osc}<150$ and $Re_n<150$.

\item The study of the temperature stratification have shown that the opposed baffles reduce its effect (more uniform temperature distribution) for the range in which it appears, $Re_{osc}<=150$ and $Re_n<200$. The stratification is negligible for higher Reynolds numbers.

\end{itemize}

For only net flow applications, due to the significant increase in the pressure drop, both configurations are suitable for applications that would require to work in laminar or transitional conditions for a smooth tube ($Re_n<4000$)  when baffles can provide a remarkable heat transfer enhancement. This is typically the case for viscous fluids that require low flow velocities to reduce the overall pressure drop, as those that can be found in chemical, petrochemical or food industries. 

The pure oscillatory flow applications are generally limited to cases where the baffles can provide a benefit in comparison to stirred tanks, e.g., a lower shear rate, that would be desirable for an algae bioreactor. 

The compound oscillatory flow applications include any process that require a high residence time (chemical reaction), what would lead to low net flow velocities and very low Reynolds numbers in a smooth tube. 

For the opposed baffles specifically, these can be recommended under net flow conditions for low net Reynolds numbers, $Re_n<150$. The same can be stated for compound flow: low oscillatory and net Reynolds numbers, $Re_{osc}<150$ and $Re_n<150$. From the point of view of the mixing performance, the observed flow patterns seem to point out a noticeable difference. Thus, it would be interesting to complement this study with techniques to measure the axial dispersion, where the opposed baffles can still be expected to produce an improvement.

\section*{Acknowledgments}

The authors gratefully acknowledge the financial support of the projects DPI2015-66493-P and PGC2018-100864-A-C22 by MCIN/AEI/ 10.13039/501100011033 and by “ERDF A way of making Europe”.

\newpage







%
\bibliographystyle{elsarticle-num} 

\bibliography{main}

\newpage

\listoffigures

\end{document}